\documentclass[twocolumn,preprintnumbers,amsmath,amssymb]{revtex4}
\usepackage{color}
\usepackage{bm, graphicx, amsmath}
\usepackage{hyperref}

\begin{document}
	
	\title{Gate-Based Circuit Designs For Quantum Adder Inspired Quantum Random Walks on Superconducting Qubits }
	\author{Daniel Koch$^{1}$$^{*}$, Michael Samodurov$^{2}$, Andrew Projansky$^{3}$, Paul M. Alsing$^{1}$}
	\affiliation{$^{1}$Air Force Research Lab, Information Directorate, Rome, NY }
	\affiliation{$^{2}$Rochester Institute of Technology, Rochester, NY}
	\affiliation{$^{3}$Hamilton College, Clinton, NY}
	\affiliation{$^{*}$Corresponding Author: dkoch.afrl@gmail.com}
	
	\begin{abstract}
		
		Quantum Random Walks, which have drawn much attention over the past few decades for their distinctly non-classical behavior,  is a promising subfield within Quantum Computing.  Theoretical framework and applications for these walks have seen many great mathematical advances, with experimental demonstrations now catching up.  In this study, we examine the viability of implementing Coin Quantum Random Walks using a Quantum Adder based Shift Operator, with quantum circuit designs specifically for superconducting qubits.  We focus on the strengths and weaknesses of these walks, particularly circuit depth, gate count, connectivity requirements, and scalability.  We propose and analyze a novel approach to implementing boundary conditions for these walks, demonstrating the technique explicitly in one and two dimensions.  And finally, we present several fidelity results from running our circuits on IBM's quantum volume 32 `Toronto' chip, showcasing the extent to which these NISQ devices can currently handle quantum walks.

	\end{abstract}
	
	\maketitle
	
	\section{Introduction}
	
	Quantum random walks are the quantum mechanical analog to classical random walks, with nearly three decades of theoretical research demonstrating their highly non-classical behaviors \cite{aharonov,childs,aharonov2,moore}.  Quantum walks quickly rose in popularity due their potential for algorithmic speedups \cite{childs2,kempe,ambainis}, as well as their mathematical flexibility to tackle a wide range of problems \cite{ambainis2,magniez,kendon,childs3}.  For nearly a decade after their inception, quantum walks remained only a mathematical framework, until the first experimental realizations of these walks came to fruition in the early 2000's.  Most notably, the first quantum technologies to demonstrate these walks were through magnetic resonance \cite{du} , photons \cite{zhang,perets,broome}, trapped ions / neutral atoms \cite{karski,xue,zahringer}, and finally superconducting qubits \cite{yan}.
	
	Similar to other quantum algorithms, it is still unclear as to which qubit technology is the most natural fit for quantum walks.  Photon based qubits offer the potential for reaching quantum walks with many steps and low decoherence \cite{su,sephton}, while trapped ions / atoms continue to show great promise for achieving walks in all sorts of new material forms \cite{xie,tamura}.  To help further this important question for quantum walks, this study focuses on the viability of implementing these walks onto superconducting qubit platforms with limited connectivity.  In particular, we analyze the strengths and weaknesses of one specific approach to designing quantum circuits, based on the Quantum Adder Algorithm \cite{qadd}.
	
	Much like the Quantum Fourier Transformation \cite{qft} (QFT), which is also at the core of the walks shown in this study, quantum random walks can be viewed in the broader sense as a quantum subroutine, fulfilling a critical role in larger quantum algorithms \cite{ven}, possibly hybrid approaches \cite{koch_scatter_1,koch_scatter_2}.  To this end, we identify standardized circuit design techniques for quantum walks on superconducting qubits as an important milestone.  Even if the quantum compositions aren't optimal, having circuit designs which can mathematically produce walks that match theoretical studies is an important first step.  And as future studies continue to refine, optimize, and improve upon these preliminary quantum walk circuits, these advances in turn improve the viability of all quantum walk based algorithms.  Our hope is that this study will lay down fundamental groundwork for implementing quantum walks, from which future research can build upon, or even inspire different techniques completely to overcome / circumvent the weaknesses of the approaches showcased here.
	
	And finally, in addition to providing several architecture-independent circuit techniques, we also demonstrate how one can optimize these walks for quantum chips with limited connectivity, adding to the steadily growing literature of superconducting based quantum random walks \cite{balu,aciate} performed on publicly available computers.  Similar to a previous study of ours \cite{koch}, we analyze the current viability of one of IBM's leading architectures \cite{ibmq_tor} for hosting quantum walks.  We present experimental results which showcase fidelity rates for core quantum walk ingredients, including QFT and QFT$^{\dagger}$.

	\subsection{Layout}
	
	The layout of this paper is as follows: section II outlines the core mathematical structure present in all walks throughout the study.  In section III, we present circuit designs for implementing a 3-qubit quantum walk tailored to IBM's qubit connectivity, as well as experimental results for QFT and Shift operations.  In section IV we introduce our circuit design for implementing boundary conditions, showcasing how the presence of boundaries can alter quantum walk probability distributions.  Section V outlines the scalability of these walks and boundary conditions into higher dimensions.  And finally, Section VI concludes the study with significant highlights throughout the paper, as well as avenues for future research.
	
	\section{Coin Quantum Walks}
	
	Throughout this paper, we will frequently use the term `quantum walk', specifically referring to the mathematical formalism of  quantum random walks driven by Coin and Shift Operators.  Classically, the roles of these two operations have direct analogies to a random walk on a discrete graph structure.  The Coin Operator mimics the process of randomly selecting an adjacent node according to some probability distribution, while the Shift Operator is analogous to physically moving the walker to the selected node.  Thus, one application of the Coin and Shift Operators in succession is the  quantum equivalent to one 'step' of a classical random walk.  However, the effects of these two operations are performed on quantum superposition states, which ultimately give rise to non-classical effects.
	
	\subsection{Mathematical Description}
	
	As with classical walks, the goal of any quantum random walk is to represent some graph structure composed of nodes and connections.  In this study, we choose to use the binary number interpretation of each quantum state as the representation of node locations.  Thus, a quantum walk composed of $N$ qubits can describe $2^N$ nodes.  However, in order for amplitudes to propagate between node connections, at least one additional qubit is needed, which is often referred to as the `coin qubit.'  The combined qubit system is then best described as being split into two subsystems, as shown in equation \ref{Eqn: QRW States}.  Quantum states created from the first system of qubits ($|N\rangle$) represent node locations on the graph, referred to as node states, while qubits in the second system ($|C\rangle$) make the various coin states.
	
	\begin{eqnarray}                 
		|\Psi\rangle \hspace{.1cm} = \hspace{.1cm} |N\rangle \otimes |C\rangle
		\label{Eqn: QRW States}
	\end{eqnarray} 
	
	The separation of qubits into node and coin subsystems is motivated by their respective roles for the Coin and Shift Operations.  More specifically, because qubits in the node system $|N\rangle$ are strictly tied to an intended graph representation, the effect of the Shift Operator is responsible for moving amplitudes around between these qubit states, based on the possible coin states of $|C\rangle$.   For example, shown below in equation \ref{Eqn: Shift Operator} is one possible mathematical formalism for a one-dimensional Shift Operator, driven by the coin states $|0\rangle$ and $|1\rangle$.  Similarly, the Coin Operator is designed to only act on the coin qubit subsystem, causing interference effects between coin states which in turn alter the amplitudes which get shifted.
	
	\begin{eqnarray}                 
		\textrm{U}_S &=& \sum_{n} |n+1\rangle\langle n|  \otimes |0\rangle\langle0|  +  |n-1\rangle\langle n| \otimes |1\rangle\langle1| 
		\label{Eqn: Shift Operator}
	\end{eqnarray} 
	
	The Shift Operator U$_S$ shown in equation \ref{Eqn: Shift Operator} above represents the mathematical structure from which all one-dimensional walks in this study will implement, whereby the coin state $|0\rangle$ causes a $+1$ shift, and similarly $-1$ for $|1\rangle$. However, it is important to note that this choice in Shift Operator is just one of many possible implementations.  In general, one has the freedom to conduct any arbitrarily complicated Shift Operator (and Coin Operator), which in principle can cause shifts by any numerical amounts, based on any set of orthogonal coin states (including higher dimensional coin states).  Such possible implementations will become more clear after the following two subsections.
	
	Contrary to the Shift Operator, whose primary role is to ultimately reflect the connectivity of a chosen graph structure, the Coin Operator can be thought of as the driving force behind a quantum walk.  Because the Shift Operator does not cause any interference effects, it falls onto the Coin Operator to cause amplitude mixing at each step.  As depicted in equation \ref{Eqn: Shift Operator}, an important quality of quantum walks is the ability to shift from a given node state $|n\rangle$ into multiple directions simultaneously, based on the amplitudes resting on the various coin states.  This feature is largely what gives quantum random walks their edge over classical counterparts, which are limited to definite probabilistic movements.  Thus, by utilizing the interference effects one can get from various U$_C$ Coin Operators, quantum walks can propagate through graph structures much quicker, and in highly non-classical ways.
	
	\subsection{Quantum Adder Shift Operator}
	
	Achieving a quantum Shift Operator like shown in equation \ref{Eqn: Shift Operator} is not unique.  Efficient methods for Shift Operator implementations is still a topic still open to future research \cite{chiang}, across all qubit hardwares.  For this study however, we focus exclusively on the viability of a Quantum Adder \cite{qadd,tutorials} based approach for superconducting qubits, so that its strengths and weaknesses can be further documented for the advancement of the overall field.  The Quantum Adder algorithm is a phase based approach to performing modulo arithmetic operations on quantum states.  This technique is particularly well suited for a coin driven quantum walk because it requires control gates, which naturally fits the $|n\pm 1\rangle \langle n | \otimes |c\rangle \langle c |$ mathematical structure of coin state based shifts.
	
	In order to use the Quantum Adder algorithm as a Shift Operator however, a necessary condition is the transformation of the node states $|N\rangle$ via the Quantum Fourier Transformation (QFT) \cite{qft}, as well as an inverse transformation (QFT$^{\dagger}$) in order to see the arithmetic result.  In between the two transformations, the gate composition of the QADD (Quantum Adder) operator is quite straightforward, as shown below in figure \ref{Fig: QADD}.  One can think of the QADD operator is adding/subtracting in `phase space' (between QFT and QFT$^{\dagger}$), ultimately achieving the desired modulo arithmetic operation when transformed back to `node space' (outside of the Fourier Transformations).
	
	\begin{figure}[h]                     
		\centering
		\includegraphics[scale=.4]{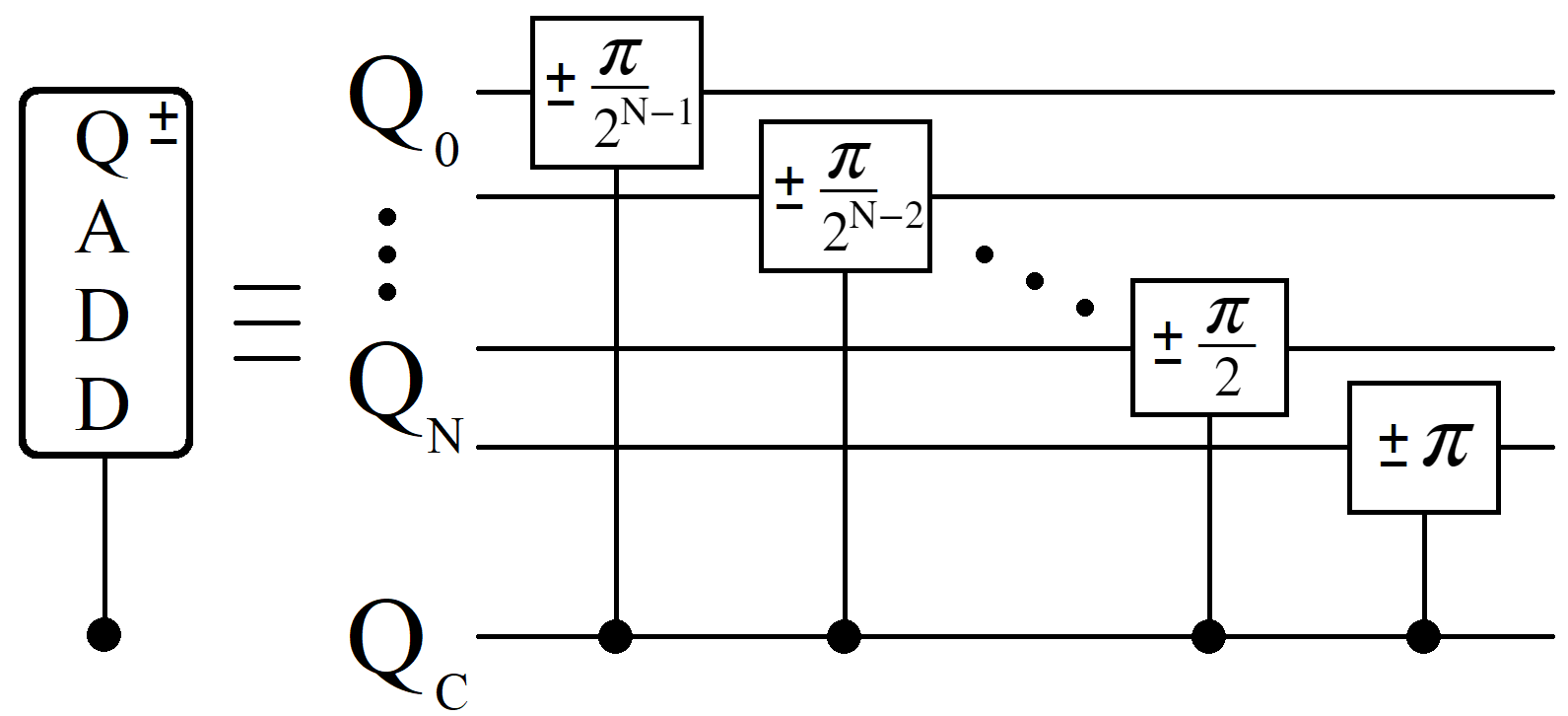}
		\caption{Quantum circuit description for the QADD operator as used throughout this paper, for the general N-qubit case, achieving a modulo arithmetic operation of $\pm 1$ on the binary node state $|\textrm{Q}_0 \textrm{Q}_1 ... \textrm{Q}_N \rangle$.}
		\label{Fig: QADD}
	\end{figure}
	
\begin{figure}[h]                     
		\centering
		\includegraphics[scale=.4]{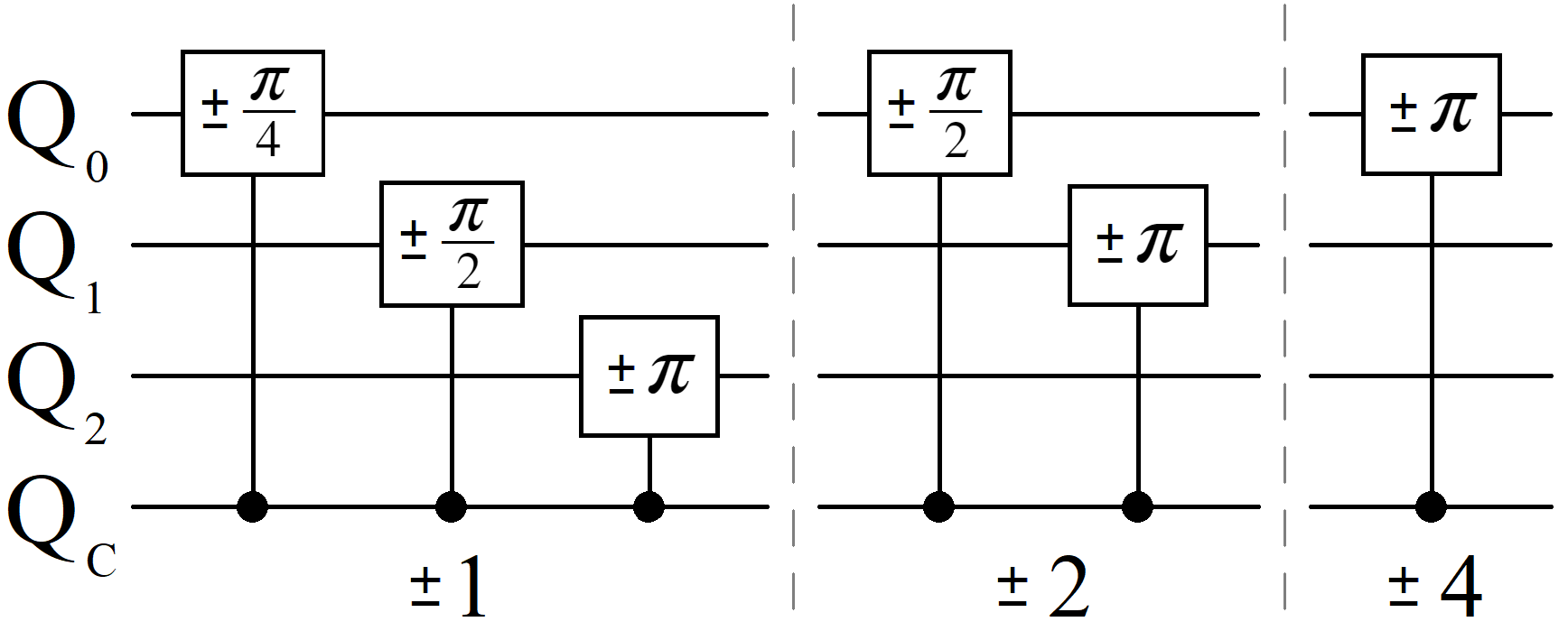}
		\caption{Three quantum circuits demonstrating how one can add/subtract various powers of $2^N$. (left) The full QADD operation for the 3-qubit case, achieving an arithmetic operation of $\pm 1$.  (center \& right) Lower order QADD operations applied to a 3-qubit system, resulting in arithmetic operations of $\pm 2$ and $\pm 4$ respectively.   }
		\label{Fig: QADD3}
	\end{figure}

	Following from the way in which the qubits are laid out in figure \ref{Fig: QADD}, the resulting QADD operator can be described as a modulo arithmetic operation acting on the  binary node state $|\textrm{Q}_0 \textrm{Q}_1 ... \textrm{Q}_N \rangle$, where Q$_N$ is the least significant bit. Additionally, operating on fewer than the full $N$ number qubits can be used to achieve addition/subtraction operations of different $2^N$ values, as shown in figure \ref{Fig: QADD3}.  In section V we show how this trick can be used to achieve shift operations on higher dimensional walks.
	
	Once again, the control-$R_{\phi}$ gates in figures \ref{Fig: QADD} and \ref{Fig: QADD3} $\textit{only}$ achieve the desired arithmetic operations when used in between QFT and QFT$^{\dagger}$ operations acting on the node qubits. Standard 3-qubit quantum circuit compositions for QFT and QFT$^{\dagger}$ can be seen in figure \ref{Apx: qftc} in the appendix.  This requirement of QFT and QFT$^{\dagger}$ for achieving U$_S$ can be viewed as the biggest technological / design hurdle for implementing this technique on NISQ \cite{nisq} hardware.  Beyond simply their gate and connectivity requirements, they create a division within the algorithm whereby the state of the node qubits should be interpreted as either node space $|N\rangle$, or phase space $|\tilde{N}\rangle$.  But only measurements performed while in node space can be interpreted as locations on a graph.  We discuss the challenge of this division in the coming sections, and its impact when introducing more complicated features into the walks such as boundaries.
	
	\subsection{Core Quantum Walk}
	
	To begin our analysis of the QADD operator based approach to quantum walks on superconducting qubits, we start with the simplest case: a one-dimensional walk on a circle.  Figure \ref{Fig: Core Walk} below shows the core quantum circuit for performing an $M$-step walk on a circle.  Separated in between the gray dashed lines are the Coin and Shift Operators, repeated $M$ times, surrounded by an initial and final QFT and QFT$^{\dagger}$.
	
	\begin{figure}[h]                     
		\centering
		\includegraphics[scale=.4]{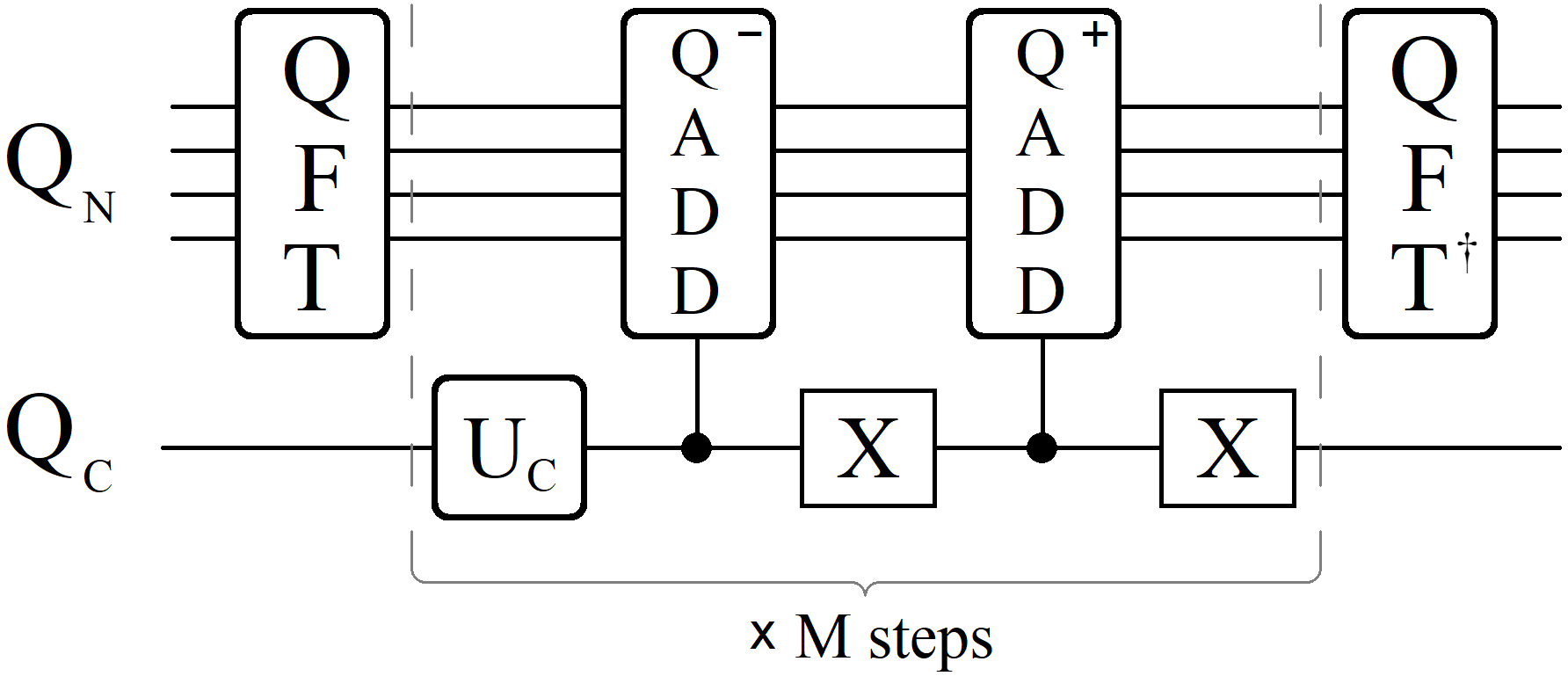}
		\caption{Quantum circuit for a 1-dimensional Quantum Random Walk with periodic boundaries, achieving the U$_S$ operator outlined in equation \ref{Eqn: Shift Operator}, for any single qubit U$_C$ Coin Operator.}
		\label{Fig: Core Walk}
	\end{figure}
	
	The circuit in figure \ref{Fig: Core Walk} represents the core structure for any quantum walk throughout this study, from which more complex walks can be further constructed.  The QADD$^{-}$ operation shifts all node states with a tensored coin state $|1\rangle$ by $-1$, while the QADD$^{+}$ operator, in combination with X gates to flip the state of the coin qubit, is responsible for shifting the remaining states by $+1$.   Because this circuit has no means of enforcing boundary conditions, the effect of QADD$^{-}$ and QADD$^{+}$ on the states $|0\rangle \otimes |1\rangle$ and $|2^N - 1\rangle \otimes |0\rangle$ respectively cause shifts which represent a graph where nodes $0$ and $2^N - 1$ are connected, i.e. a circle.
	
\begin{figure}[h]                     
		\centering
		\includegraphics[scale=.4]{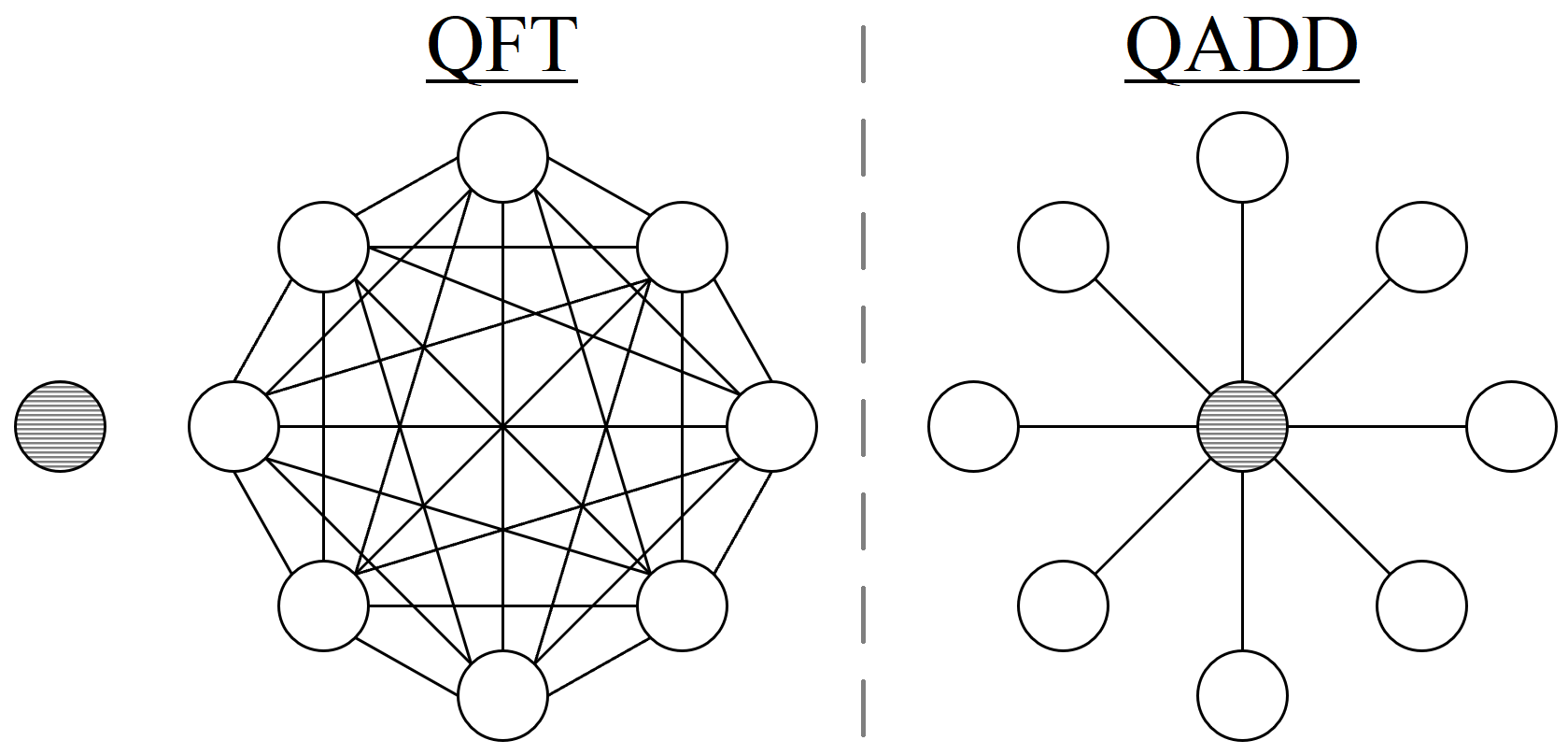}
		\caption{ Qubit connectivity requirements for the QFT (left) and QADD (right) operations, for a one dimensional 8-qubit quantum walk.  Node qubits are represented by white circles, while the coin qubit is shown with gray stripes.}
		\label{Fig: Connectivity}
	\end{figure}

	In analyzing the simplest one-dimensional walk, there are two important features of figure \ref{Fig: Core Walk} which are already worth noting.  Firstly, the number of steps ($M$) for the quantum walk is independent of QFT and QFT$^{\dagger}$.  Or stated in another way, the entire quantum walk can be achieved with a single transformation in and out of phase space, which is the optimal case.  We note this here because later this condition will change once boundaries are introduced.  Secondly, despite being the simplest case, the circuit shown above requires full inter-connectivity between all $N+1$ qubits.  The standard circuit implementations for $N$-dimensional QFT and QFT$^{\dagger}$ operations require full connectivity between the node qubits, while an additional connection is required between each node qubit and the coin qubit for the QADD$^{\pm}$ operations.   However, this full connectivity requirement can be relaxed based on which operation is being performed, as illustrated in figure \ref{Fig: Connectivity}.
	
	As demonstrated in figure \ref{Fig: Connectivity}, the two core operations for implementing U$_S$ on any quantum walk (QFT and QADD) require vastly different connectivity between qubits.  The QFT and QFT$^{\dagger}$ operations impose the stricter of the two connectivities, while the QADD operations only require connectivity between the coin qubit and each of the node qubits individually.  In the next section we demonstrate how one can take advantage of these two differing connectivities to execute quantum walks on NISQ devices with limited connectivity. However, it is worth noting that future generations of NISQ architectures may soon deliver on all-to-all connectivity between qubits \cite{IonQ}, greatly improving the outlook of quantum walks viability.

	\section{Implementing a 3-Qubit Walk}
	
	Having outlined the core components of any quantum walk for this study in the previous section, we now turn our attention to the realization of these walks on NISQ era devices.  In particular, we focus our efforts on designing and analyzing quantum circuits which follow IBM's `heavy hexagonal' \cite{hh} connectivity for superconducting qubits.  Since this design is anticipated to be prominent throughout their future quantum architectures in preparation for quantum error correcting, we consider the strengths and challenges of implementing quantum walks to this connectivity (without error correcting).
	
	\subsection{QFT and QFT$^{\dagger}$ Circuits}
	
	In designing quantum circuits tailored to the heavy hexagonal geometry, one finds that the highest number of qubit connections belonging to a single qubit is three, shown below in figure \ref{Fig: 3QC}.  This connectivity naturally matches the ideal geometry for a 3-qubit QADD operation, with the coin qubit centralized between three neighboring node qubits.  However, while ideal for the QADD operator, this connectivity is also naturally the $\textit{worst}$ possible geometry for performing a QFT operation between the outer nodes (zero connections between node qubits).
	
	\begin{figure}[h]                     
		\centering
		\includegraphics[scale=.3]{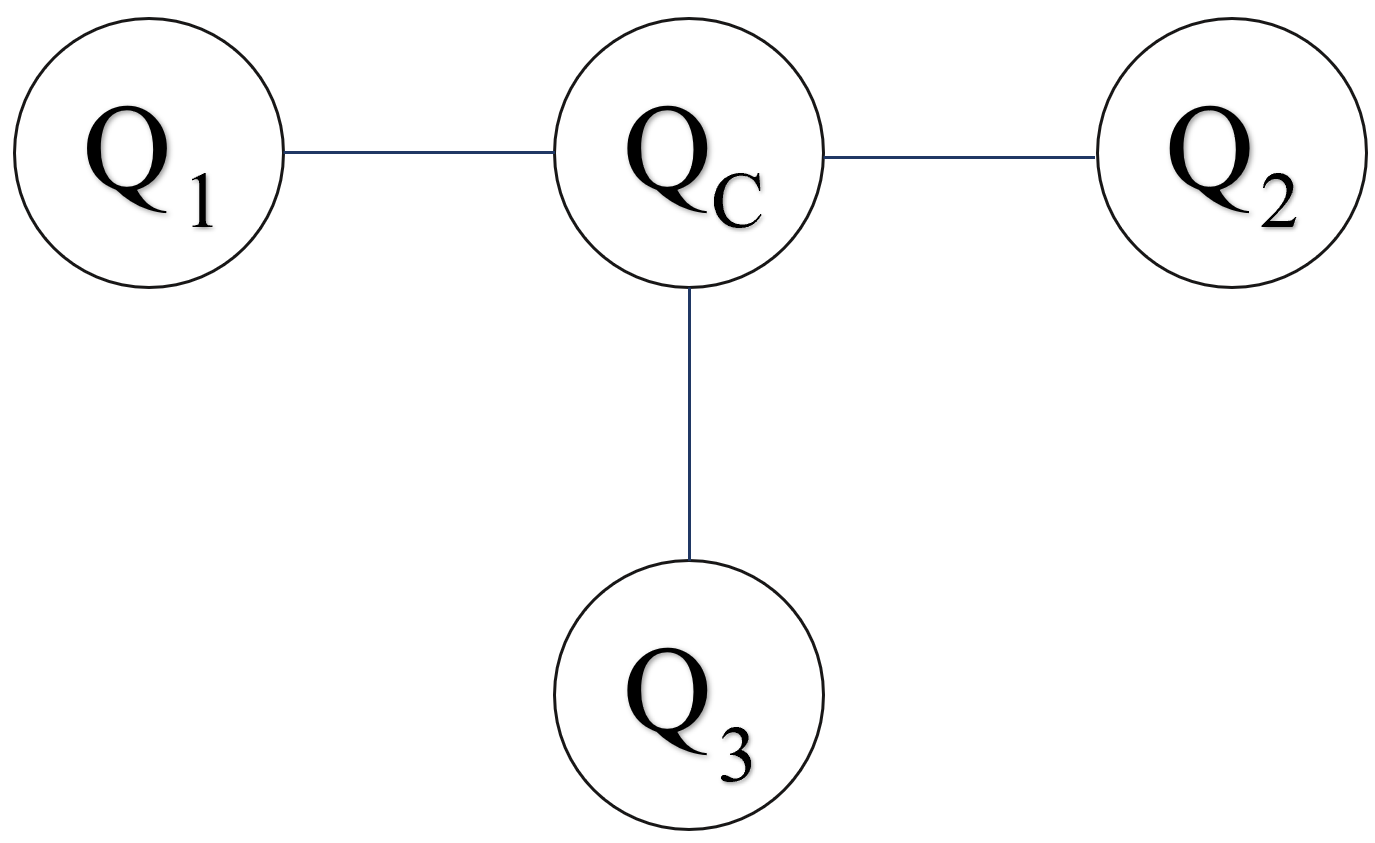}
		\caption{ The highest qubit connectivity found on IBM's heavy hexagonal \cite{hh} chip designs.  Due to the connectivity requirement of the QADD operator, the connectivity shown here is ideal for implementing the Shift Operator for a 3-qubit quantum walk in one dimension. }
		\label{Fig: 3QC}
	\end{figure}
	
\begin{figure}[h]                     
		\centering
		\includegraphics[scale=.4]{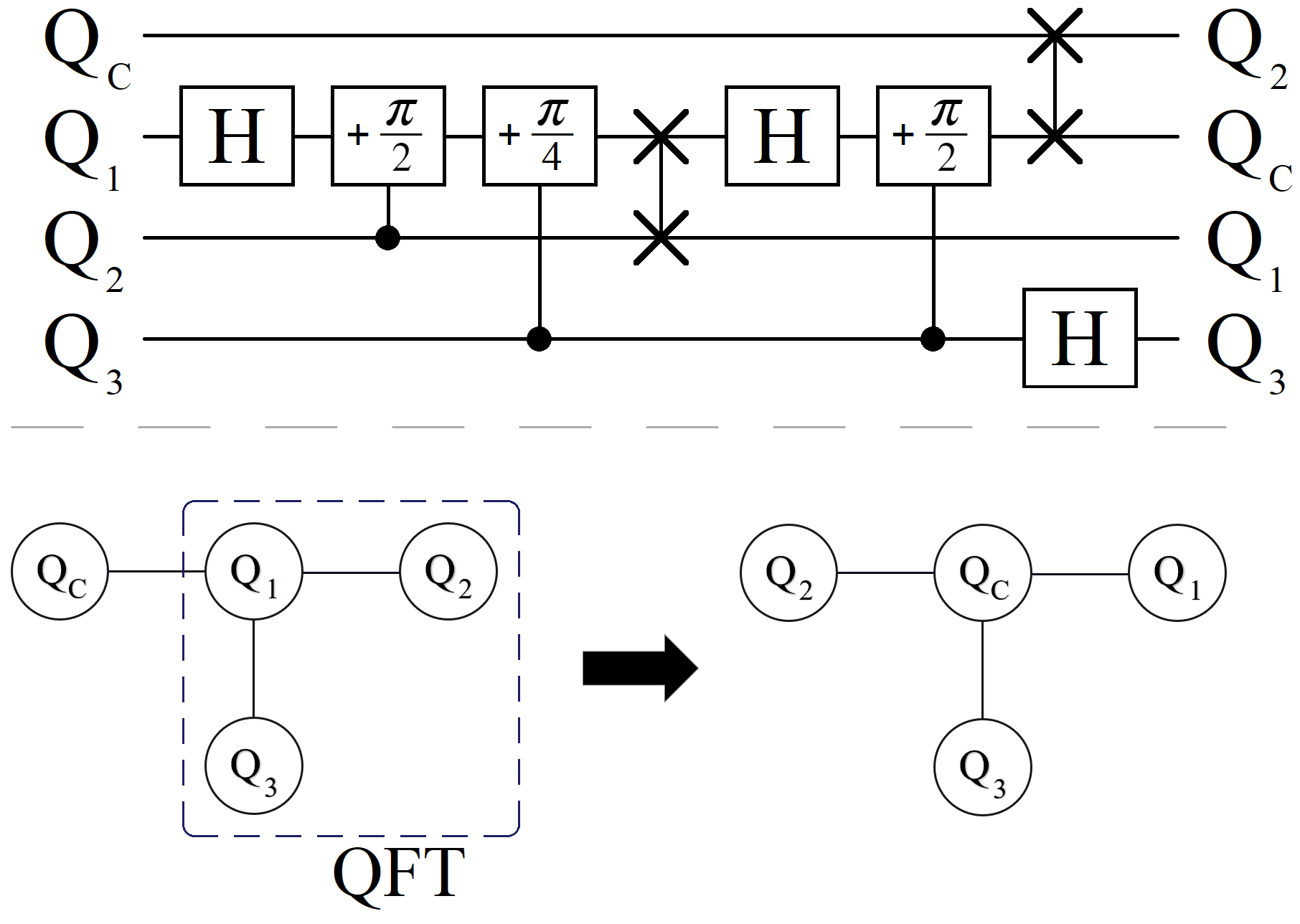}
		\caption{Quantum circuit for an initial 3-qubit QFT following the qubit connectivity outlined in figure \ref{Fig: 3QC} (also shown at the bottom of the figure).  The design of the circuit is such that the number of SWAP gates is minimal, while leaving the coin qubit centralized for future QADD operations.  }
		\label{Fig: HH QFT}
	\end{figure}

	Optimal implementation of quantum operations such as QFT with limited qubit connectivity is a NISQ era specific challenge in quantum computing.  In a past study, we observed the fidelity rates between several such circuit designs \cite{koch}, including the geometry shown in figure \ref{Fig: 3QC}.  Here however, the implementation of QFT and QFT$^{\dagger}$ on this qubit geometry is more difficult, as the central qubit (the coin qubit) cannot be treated as simply an ancilla for supplementing missing connections, but rather must have its quantum state always preserved.  To this end, we propose an adapted QFT circuit, shown in figure \ref{Fig: HH QFT}, which performs a QFT between the three node qubits, while simultaneously preparing the system for future QADD operations.
	
	Illustrated by the two diagrams at the bottom of figure \ref{Fig: HH QFT}, the goal of the quantum circuit is use the central qubit's connectivity as advantageously as possible for both the QFT and QADD operations.  By utilizing SWAP gates, which trade quantum states held between two qubits, the physical locations of the node and coin qubits can be swapped throughout the algorithm.  As indicated by the dashed border in the figure, the circuit begins by treating the three linearly connected qubits as node qubits for the QFT operation.  With this starting configuration, two out of the three necessary connections for QFT are present, requiring only one SWAP gate for the final connection needed between qubits Q$_2$ and Q$_3$. 
	
	Following all of the gates necessary for the initial QFT, qubits Q$_1$ and Q$_2$ have traded physical locations, leaving Q$_2$ as the central qubit in the geometry.  Thus, in order to produce the connectivity needed for the following QADD operation(s), a second SWAP gate is used between Q$_2$ and Q$_C$, resulting in the final configuration shown in figure \ref{Fig: HH QFT} (bottom right).  In total, this circuit design leaves three out of four qubits in physical locations which differ from their starting location.  Qubits Q$_C$, Q$_1$, and Q$_2$ all change positions on the physical chip, but their quantum states successfully hold the effect of a 3-qubit QFT between qubits Q$_1$, Q$_2$, and Q$_3$ (besides swapping locations, the state of Q$_C$ never changes).  So long as one is aware of these changes however, the only consequence is that one must alter the respective control-R$_{\phi}$ gates for the following QADD operation(s), properly applying the correct phases to each physical qubit.  And finally, after $M$ steps of the quantum walk, the QFT$^{\dagger}$ operation will result in the return of each qubit to its starting physical location.
	
	\subsection{QFT Experimental Results}
	
	In order to better understand the viability of implementing these quantum walks on real hardware, as well as scalability, in this subsection and the next we present experimental results gathered from one of IBM's leading quantum volume chips: Toronto \cite{ibmq_tor,qv}.  Each experiment was performed on the various 4-qubit groupings found across the chip, matching the connectivity of figure \ref{Fig: 3QC}.  For the first experiment, we test the quantum circuit laid out in figure \ref{Fig: HH QFT}, with the full quantum score given by figure \ref{Apx: IBMQ QFT} in the appendix.  The results shown below in figure \ref{Fig: QFT Data} demonstrate fidelity rates found from applying QFT followed by QFT$^{\dagger}$  to various 3-qubit initial states, as defined below in equation \ref{Eqn: QFT Fidelity}.
	
	\begin{eqnarray}                 
		f  \equiv \langle \Psi_i | \textrm{QFT}^{\dagger} \textrm{QFT} | \Psi_i \rangle \otimes  \langle 0 | \textrm{Q}_C \rangle
		\label{Eqn: QFT Fidelity}
	\end{eqnarray} 
	
	\begin{figure}[h]                     
		\centering
		\includegraphics[scale=.50]{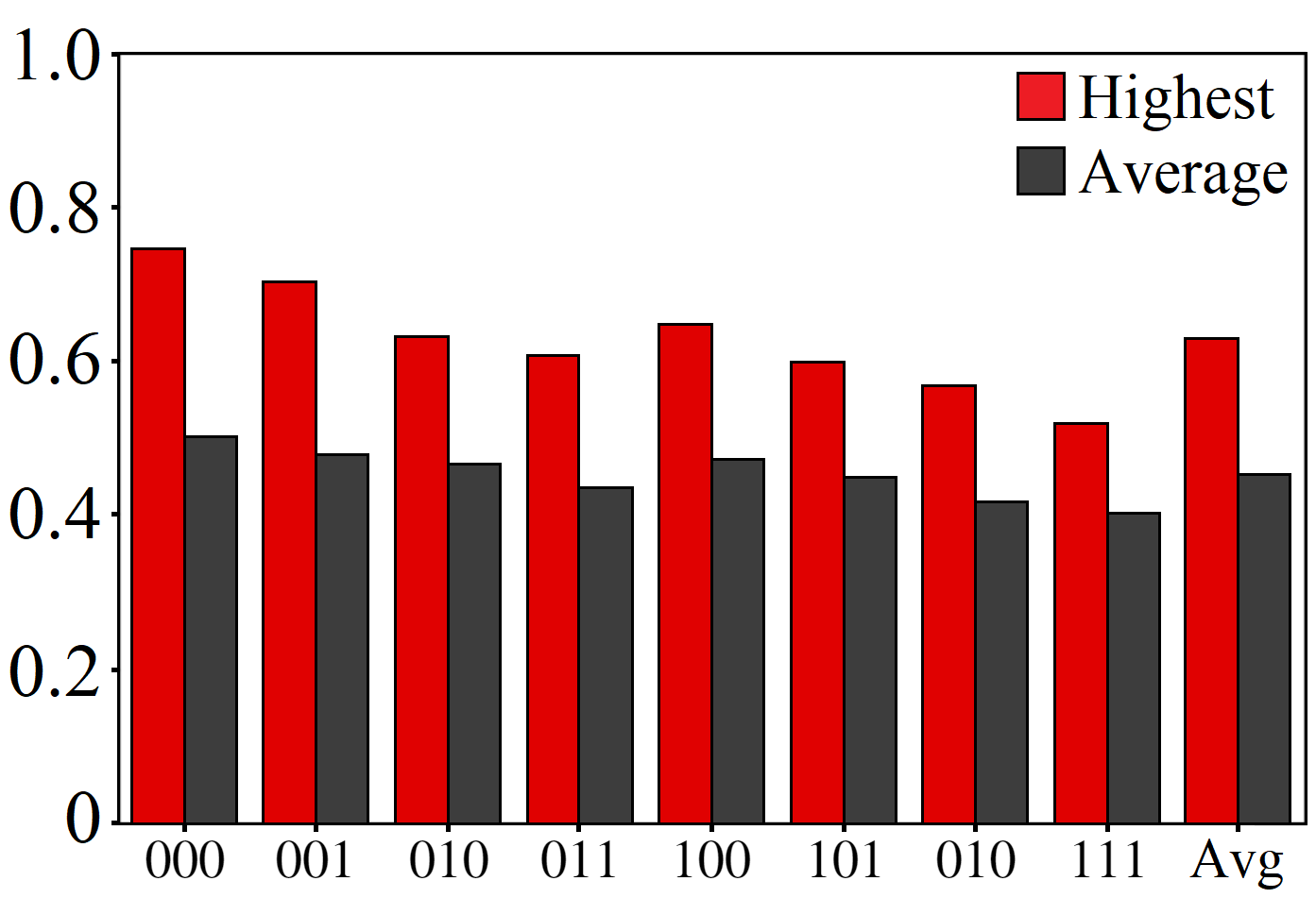}
		\caption{Fidelity rates (equation \ref{Eqn: QFT Fidelity}) for the eight possible initial states after applying sequential QFT and QFT$^{\dagger}$ (figure \ref{Fig: HH QFT}) operations, tested across all eight 4-qubit combinations (figure \ref{Fig: 3QC}) present on the Toronto architecture. (red) The highest fidelity rates found for a single 4-qubit combination. (dark gray) Average fidelities found across all qubit combinations.  Each initial state was tested 16,000 times per qubit combination, for a total of 1,024,000 experimental runs. }
		\label{Fig: QFT Data}
	\end{figure}
	
	Figure \ref{Fig: QFT Data} demonstrates the percentage of experiments whereby sequential QFT \& QFT$^{\dagger}$ operations returned each of the qubits back to their respective initial states, additionally requiring the coin qubit be in the $|0\rangle$ state (initialized in the $|0\rangle$ state).  The results show the average fidelities found for each initial state, across all of the different 4-qubit configurations on the Toronto chip (dark gray), as well as the configuration with the highest overall fidelity (red).  Both bar plots indicate a slightly decreasing trend in the direction of node states containing more $|1\rangle$ state qubits (with $|111\rangle$ being the lowest).  This result is understandable, given that these states are subject to more T$_1$ errors \cite{T1_1,T1_2,koch} (all of the initial states should in principle be subject to the same T$_2$ errors \cite{T2_1,T2_2} after the first QFT).
	
	The results of figure \ref{Fig: QFT Data} can be interpreted as baseline fidelities from which more complex quantum walks can be built on.  This is because any Quantum Adder based approach to implementing quantum walks $\textit{must}$ include at least one pair of QFT and QFT${^{\dagger}}$ operations for achieving U$_S$.  Naturally then, the next minimum to any quantum walk is to include at least one QADD operation, which we now show in the coming subsection.

	\subsection{Shift Operator Results}

\begin{figure}[h]                     
		\centering
		\includegraphics[scale=.50]{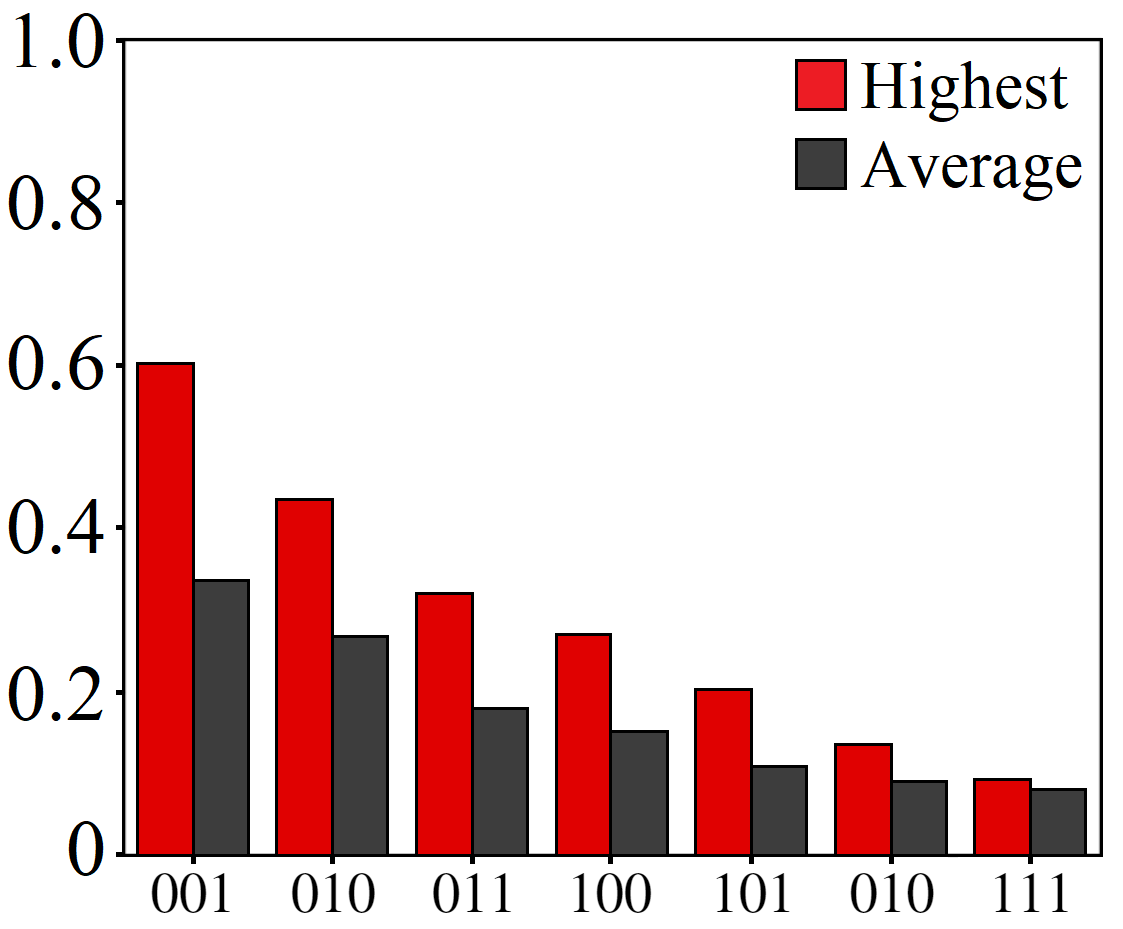}
		\caption{Fidelity rates (equation \ref{Eqn: Shift Fidelity}) for up to 7 shift operations, tested across all eight 4-qubit combinations present on the Toronto architecture. (red) The highest fidelity rates found for a single 4-qubit combination.  (dark gray) Average fidelity across all qubit combinations.  Each shift operation order N was tested 16,000 times per qubit combination, for a total of 896,000 experimental runs. }
		\label{Fig: Shift Data}
	\end{figure}

	The results shown in figure \ref{Fig: Shift Data} below demonstrate the complete Quantum Adder Shift Operator, with the goal of achieving amplitude shifts of up to 7 nodes.  Each experiment starts in the initial state $|000\rangle$, and aims to achieve shifts to all other basis states through repetitions of the QADD operator.  The fidelity for each experiment is defined below in equation \ref{Eqn: Shift Fidelity}.  One again, average fidelity results are derived from testing all possible 4-qubit combinations across the Toronto architecture, as well as the highest fidelities found for a single combination.
	
	\begin{eqnarray}                 
		f  \equiv \langle N| \textrm{QFT}^{\dagger} \hspace{.03cm} ( \textrm{QADD}^+ )^N \hspace{.03cm}  \textrm{QFT} | 000 \rangle \otimes  \langle 0 | \textrm{Q}_C \rangle
		\label{Eqn: Shift Fidelity}
	\end{eqnarray} 
	
	As expected, increasing the number of QADD operations between the two Fourier Transformations results in lower fidelity rates.  Each application of QADD$^{+}$ for three qubits is approximately 6 CNOT and 9 phase gates (see appendix figure \ref{Apx: IBMQ Shift}), which quickly increase the circuit's total length and gate count per iteration. The results demonstrate that an optimistic single step walk on current NISQ hardware can yield fidelities around 50-60 \%, but anything larger than four steps and the fidelity drops below 25 \%.  Additionally, the experiments for figure \ref{Fig: Shift Data} were conducted such that the state of the coin qubit was only excited (to $|1\rangle$) during the QADD operations, and returned to $|0\rangle$ for the QFT operations in order to provide a small boost in fidelity (minimizing T$_1$ errors).  Thus, also requiring the coin qubit to preserve superposition throughout the walk would be expected to yield even smaller fidelities still.
	
	The results of figures \ref{Fig: QFT Data} and \ref{Fig: Shift Data} can be summarized as the viability of implementing one-dimensional walks on IBM's state-of-the-art qubits. Although these fidelities are currently too low for algorithmic use, we encourage the reader to keep in mind that these numbers will improve consistently with time, as subsequent generations of qubits become available.  Looking forward, the true merit to these benchmarking experiments is to continually monitor the progress of real qubits as the technology approaches the threshold for which quantum walks can provide an algorithmic speedup.

	\section{Node Boundaries}
	
	Having now seen the extent to which IBM's current superconducting qubits can handle the core elements of a one-dimensional quantum walk, the remaining focus of this study will be to introduce and analyze more complex features desired of quantum walks.  More specifically, in this section we propose a novel circuit approach for implementing boundaries in one dimension, which we then carry over to higher dimensional walks in the next and final section. 
	
	\subsection{Enforcing Boundary Conditions}
	
	In order for quantum walks to compete with classical counterparts for advantageous speedups, a necessary milestone is the ability to freely implement them onto desirable graph structures.  This includes the ability to shift in numerous directions from a single node, with probabilities specific to each node.  Mathematically, writing down unitary operators to handle such complex walks is akin to equation \ref{Eqn: Shift Operator}, but creating these walks with quantum circuits is a much more complicated problem.  
	
	In pursuit of creating quantum walks on arbitrarily complex graphs, we identify enforcing boundaries as an important first step.  For the context of this study, the term `boundary' will refer to a shift between two node states which is forbidden.  For example, in performing a quantum walk in one dimension with three qubits, a boundary between nodes 5 and 6 would mathematically mean the following two shifts are forbidden:
	
	\begin{eqnarray}                 
		\textrm{U}_S \hspace{.05cm} |101\rangle  \otimes |0\rangle  \not\rightarrow |110\rangle \otimes |0\rangle  \\
		\textrm{U}_S \hspace{.05cm} |110\rangle  \otimes |1\rangle  \not\rightarrow |101\rangle \otimes |1\rangle \nonumber
		\label{Eqn: No Shift}
	\end{eqnarray} 
	
	In order to enforce a boundary like shown above, we consider two approaches to the problem.  The first approach is to alter the U$_S$ Shift Operator, preventing the shifts shown in \ref{Eqn: No Shift} from occurring, but leaving all other shifts throughout the system unchanged.  The second approach, which we have chosen to examine more fully in this study, would be to prevent the states $|101\rangle  \otimes |0\rangle$ and $|110\rangle  \otimes |1\rangle$ from occurring within the walk by altering the Coin Operator U$_C$, leaving U$_S$ unchanged.  
	
	In regards to the first approach, it turns out that there is an unavoidable conflict which occurs when using the Quantum Adder Shift Operator as U$_S$.  In particular, creating an adapted version of U$_S$ which affects unique node states differently is made difficult by the presence of QFT and QFT$^{\dagger}$.  Because the QADD operator requires Fourier Transformations, the natural divide within the algorithm between `node space' and `phase space' directly impacts when standard higher order control operations using sequential Toffoli gates \cite{toffoli,toffoli2} are applicable or not.  Outside of the QFT's (node space), these higher order control chains through ancilla qubits can be used to target individual node states among the system's total superposition, but not so within the QFT's.  For the purpose of boundaries, this makes targeting select node states for different shift operations difficult.  Thus, a control-Shift Operator approach to boundaries would require clever use of state marking (through excited ancilla qubits) before the QFT at each step, such that select node+coin states do not shift across boundaries.
	
	\begin{eqnarray}                 
		\textrm{U} =  U_{\textrm{QFT}} \hspace{.15cm} \textrm{U}' \hspace{.15cm} U_{\textrm{QFT}^{\dagger}}
		\label{Eqn: QFT Matrix}
	\end{eqnarray}
	
	As an alternative approach, one could in theory write down the complete unitary operator U, which correctly shifts all of the states in the system according to boundary conditions, and then solving equation \ref{Eqn: QFT Matrix} above for U$^{'}$, yielding the matrix operation needed between QFT and QFT$^{\dagger}$.  However, translating U$^{'}$ into standard quantum gates , which can be an arbitrarily complex $2^N$ dimensional matrix, is the primary issue here.  While several mathematical techniques for handling this problem are known \cite{motto,daskin}, applying these decompositions to U$^{'}_S$ (the altered QADD operator) is impractical for two reasons: 1) As the size of the walk grows, the gate count for the resulting U$^{'}_S$ operation quickly becomes infeasible for any optimistic NISQ devices. 2) Implementing a boundary at any given location, or combination of locations, requires a unique U$^{'}_S$, and it is unclear if there is any  easily recognizable pattern to their corresponding gate compositions, or if a new solution is necessary for every unique graph.

	\subsection{Coin Boundaries}
	
	Instead of altering the Quantum Adder Shift Operator in order to enforce boundaries, which we leave as an open avenue for future research, here we consider a Coin Operator based approach.  Rather than strictly requiring  that certain transitions between select nodes be impossible, this approach anticipates and changes coin states for nodes with boundaries, which would otherwise violate boundary conditions when acted upon by U$_S$ with no intervention.  
	
	\begin{figure}[h]                     
		\centering
		\includegraphics[scale=.30]{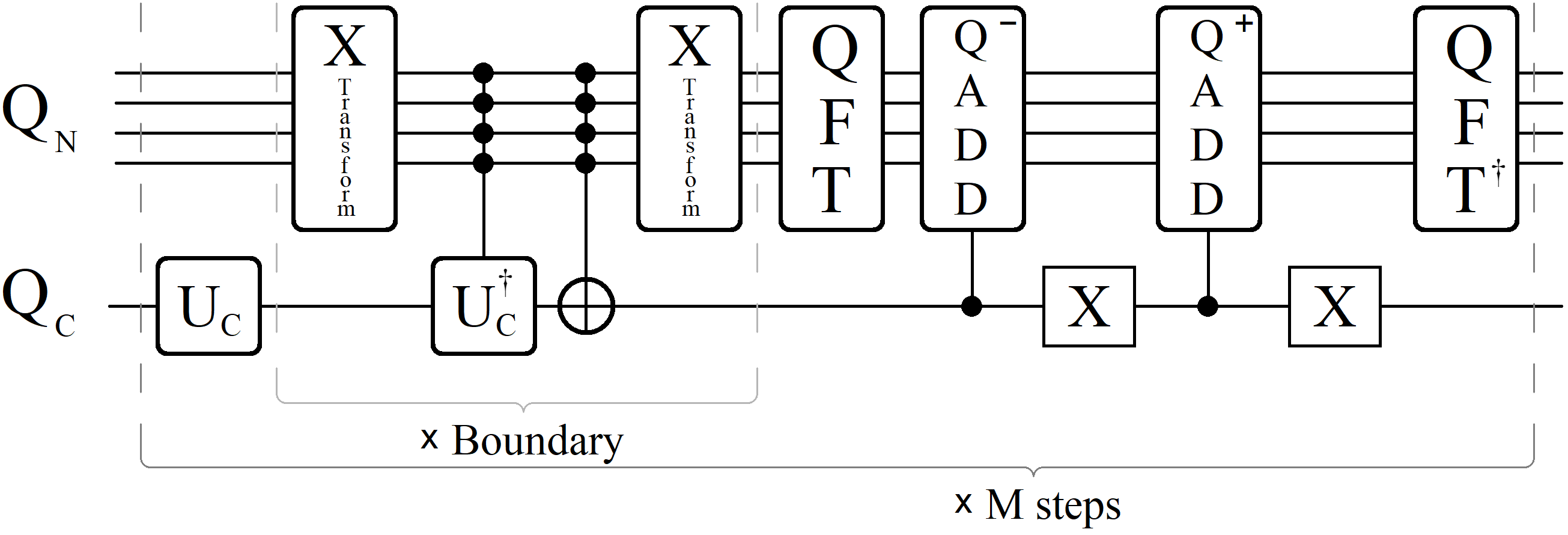}
		\caption{Quantum circuit for implementing a general coin boundary, for a quantum walk driven by a U$_C$ Coin Operator.  The 'X Transform' operator corresponds to the required single qubit X gates necessary for transforming the desired boundary node state to the state of all $|1\rangle$'s. }
		\label{Fig: Coin Boundary Walk}
	\end{figure}
	
	Figure \ref{Fig: Coin Boundary Walk} above outlines the general strategy for implementing boundaries via the Coin Operator.  To summarize, the goal is to `undo' the effect of the Coin Operator U$_C$ for specific node states (the location of the desired boundary) within the superposition, followed by a flipping of the coin state.  By canceling the effect of the Coin Operator with U$^{\dagger}_C$, this ensures that the targeted node+coin state remains unchanged from the previous operation in the circuit, which is expected to be the Shift Operation from the previous step.  Knowing this, following the control-$U^{\dagger}_C$ operation with a flip of the coin state (via the same dimensional control-X) guarantees that no amplitude shifts will cross the desired boundary during the next shift operation.  Figure \ref{Fig: CB Example} below illustrates an example of this approach, for the case of a boundary between nodes 5 and 6.
	
	\begin{figure}[h]                     
		\centering
		\includegraphics[scale=.55]{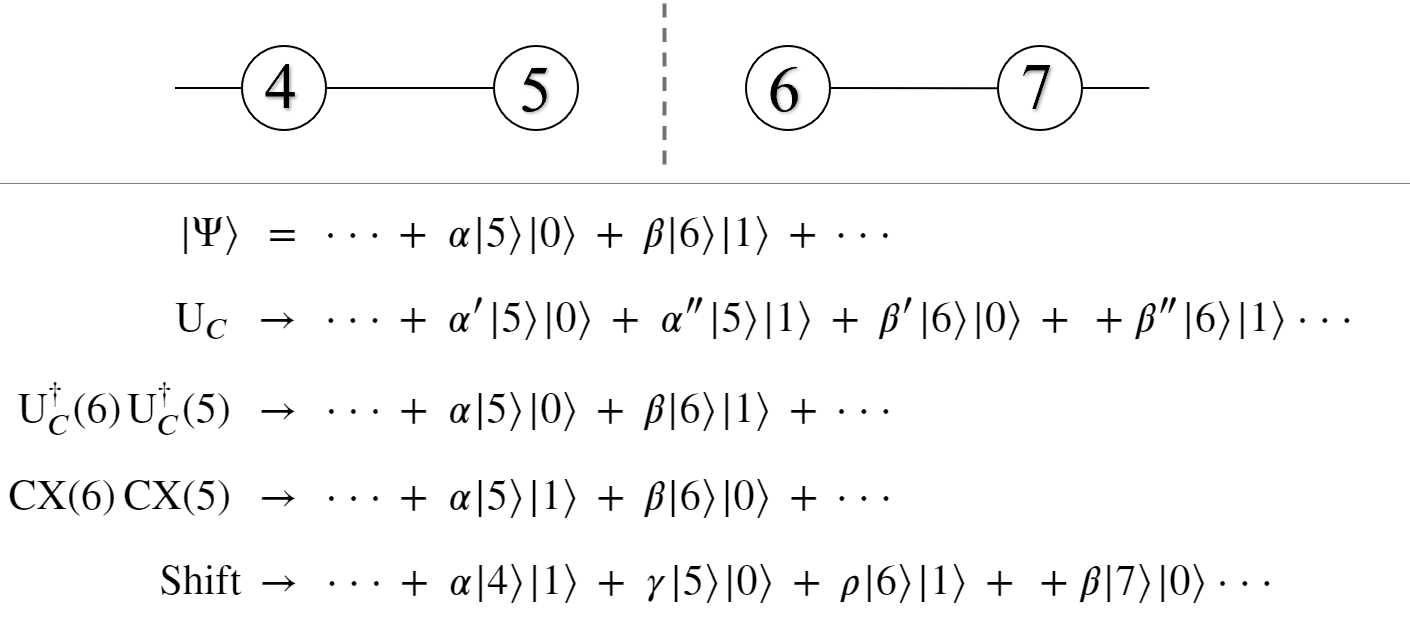}
		\caption{Step by step view of how a boundary coin operation prevents an amplitude shift between nodes 5 and 6. U$^{\dagger}_C$($N$) and CX($N$) both refer to higher order control operations which use the node state $|N\rangle$ as a control, resulting in U$^{\dagger}_C$ and X operations on the coin qubit (see figure \ref{Fig: Coin Boundary Walk}).  Note that in the final step, states $\gamma |5\rangle |0\rangle$ and $\rho |6\rangle |1\rangle$ come from shifts from nodes $4$ and $7$ respectively, as all other nodes in the system contain both coin states at every step. }
		\label{Fig: CB Example}
	\end{figure}
	
	Following the steps shown above in figure \ref{Fig: CB Example}, note that the state $|5\rangle |0\rangle$ is no longer present in the superposition just prior to the shift, which means that the transition $ |5\rangle |0\rangle \hspace{.05cm} \longrightarrow \hspace{.05cm} |6\rangle |0\rangle $ never occurs, successfully enforcing a boundary between nodes 5 and 6.  Similarly, the same effect occurs from the reverse direction, preventing the state $|6\rangle|1\rangle$ from causing a shift across the boundary.  Importantly, this technique is designed such that as the forbidden superposition states ( $ |5\rangle |0\rangle$ and $ |6\rangle |1\rangle$ in this case ) aren't present in the initialization of the walk, and U$_C$ is a single qubit operation, the circuit outlined in figure \ref{Fig: Coin Boundary Walk} ensures that these forbidden boundary coin states will $\textit{never}$ be present just prior to each step's Shift Operator.  Thus, after proper initialization of the system, the boundary conditions will hold indefinitely.  Additionally, we also note that there is a degree of freedom with which the amplitudes at either side of the boundary can be reflected back with, shown below in equation \ref{Eqn: 1D Boundary U}, which could potentially serve to improve certain quantum walk applications.
	
	\begin{eqnarray}                 
		\textrm{U}_B  \hspace{.02cm} |N\rangle \otimes |C\rangle \hspace{.05cm} \longrightarrow \hspace{.05cm} e^{i\phi}  \hspace{.02cm} |N\rangle \otimes X |C\rangle
		\label{Eqn: 1D Boundary U}
	\end{eqnarray} 
	
	\subsection{Pros \& Cons of Coin Boundaries}
	
	Another way to describe the coin boundary technique outlined in the previous subsection is to say that the effect of the Coin Operator U$_C$ happens at every node location (on every node state $|N\rangle$) except those directly adjacent to the desired boundary.  By applying U$^{\dagger}_C$ to these boundary node states, effectively canceling the Coin Operator's amplitude mixing, the state of the coin qubit for these node states is predictable, as there is only one possible direction from which amplitudes can then arrive at the node state.  However, in order for the boundary to hold throughout the walk, both sides of the boundary need to be enforced at every step.  If any amplitude is leaked across the boundary, or the system is improperly initialized, the process breaks down.
	
	By comparison to using an altered U$_S$ strategy for enforcing boundaries, the circuit depicted in figure \ref{Fig: Coin Boundary Walk} is significantly easier to implement than any U$^{'}_S$ alternative.  However, the biggest disadvantage to using coin boundaries isn't in the gate composition itself, but rather once again in QFT and QFT$^{\dagger}$.  If one compares the circuits in figures \ref{Fig: Core Walk} and \ref{Fig: Coin Boundary Walk}, the most significant difference is the fact that the QFT and QFT$^{\dagger}$ operations become included within the $M$ repeated steps, whereas previously they were independent.  Because the coin boundary technique requires control chains based from node states $|N\rangle$, the consequence is that every step requires a new QFT and QFT$^{\dagger}$ in order to implement the QADD based Shift Operator, quickly driving up gate count and circuit depth.

	\subsection{One Dimensional Bounded Walk}
	
	To illustrate the kinds of effects that can arise from using Coin Operator based boundaries, here we present a few simulated results found while studying Hadamard Coin driven one-dimensional walks with a boundary between the nodes 0 and $2^N\textrm{-}1$ (for a walk composed of N node qubits).  Each walk begins in an initial equal superposition between all node states (H$|0\rangle$ for each node qubit), with an initial coin state of $|0\rangle$.  With no boundaries, the effect of this walk gives rise to wavefunctions which propagate like waves in both directions \cite{kempe}, with the larger wave traveling to the right.  With periodic boundaries, these waves eventually collide and produce interference effects.  However, with Coin Boundaries preventing shifts between nodes $0$ and $2^N\textrm{-}1$, the effect is quite different. 
	
	\begin{figure}[h]                     
		\centering
		\includegraphics[scale=0.6]{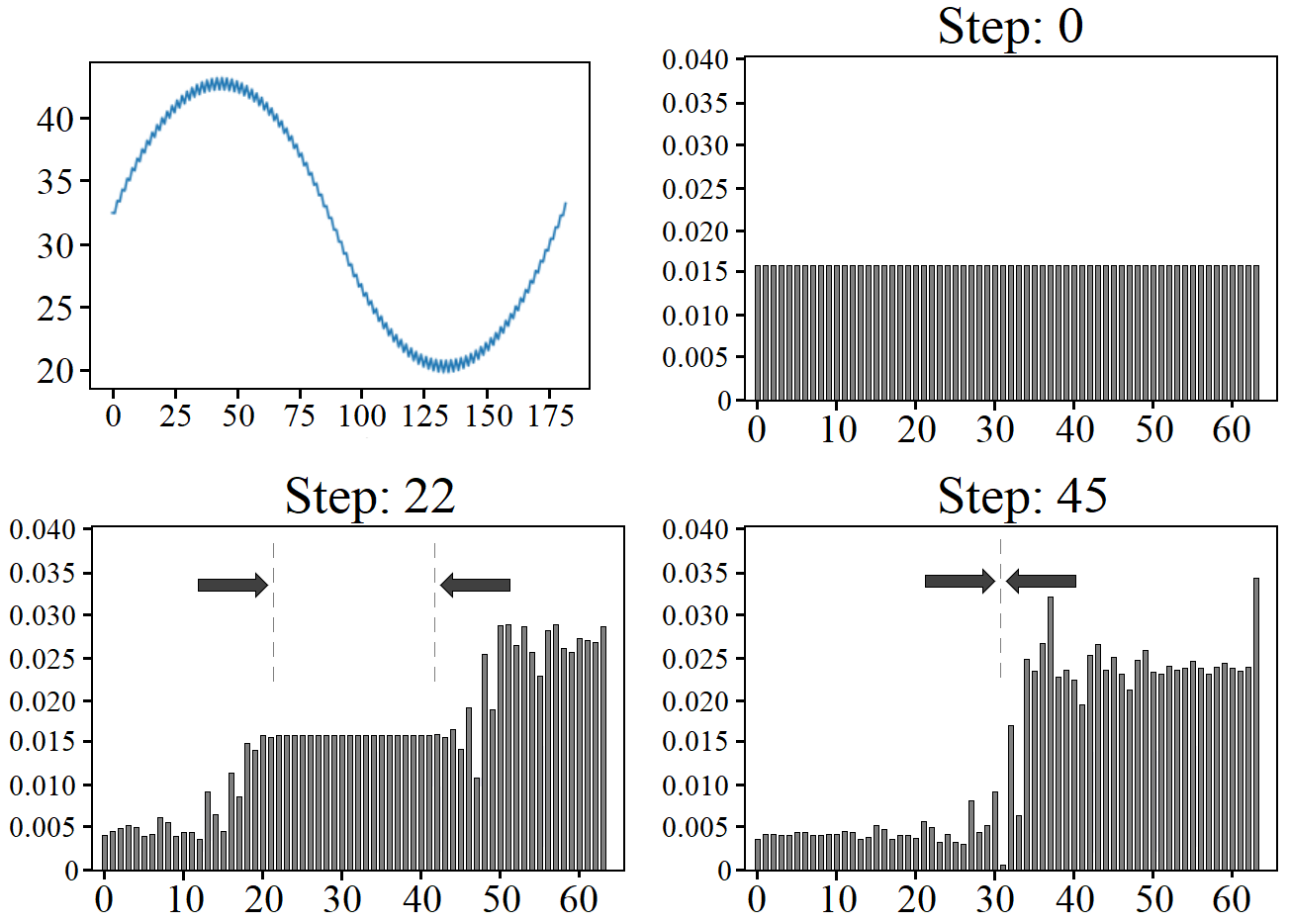}
		\caption{(top left) A plot showing average node probability as a function of steps (see equation \ref{Eqn: Node Prob}).  The remaining three figures each show how the total probability in the system is concentrated among the $2^N$ node states, showing how the effect of the Coin Boundaries propagate in from both sides.  }
		\label{Fig: CB Prob Plot}
	\end{figure}
	
	The plots shown in figure \ref{Fig: CB Prob Plot} illustrate the effect of Coin Boundaries on a one dimensional quantum walk.  Shown in the top left corner is a plot of average node probability as a function of steps in the walk, given by equation \ref{Eqn: Node Prob}, for the case of a 64 node walk created from 6 qubits.  All of the results shown in the figure were generated using IBM's Qiskit simulator, representing a noiseless quantum walk.
	
	\begin{eqnarray}                 
		\textrm{Average Node Probability} \hspace{.05cm} = \hspace{.05cm}  \sum_{n=0}^{2^N - 1}  n \hspace{.03cm}|  \langle n | \Psi \rangle |^2
		\label{Eqn: Node Prob}
	\end{eqnarray} 
	
	Accompanying the average node probability plot are three instances of the walk, showing how the total probability of the system is distributed for steps 0, 22, and 45 respectively.  At step 0, all of the probability in the system is evenly distributed across all 64 node states, with the state of the coin qubit being $|0\rangle$.  Consequently, this choice in initial coin state causes average node probability to flow to the right (towards node 63) with each step.  This effect can be understood as the consequence of having the $|0\rangle$ coin state determine positive shifts, and $|1\rangle$ for negative shifts, which leads to deconstructive interference around node 0, and similarly constructive interference around node 63. 
	
\begin{figure}[h]                     
		\centering
		\includegraphics[scale=0.48]{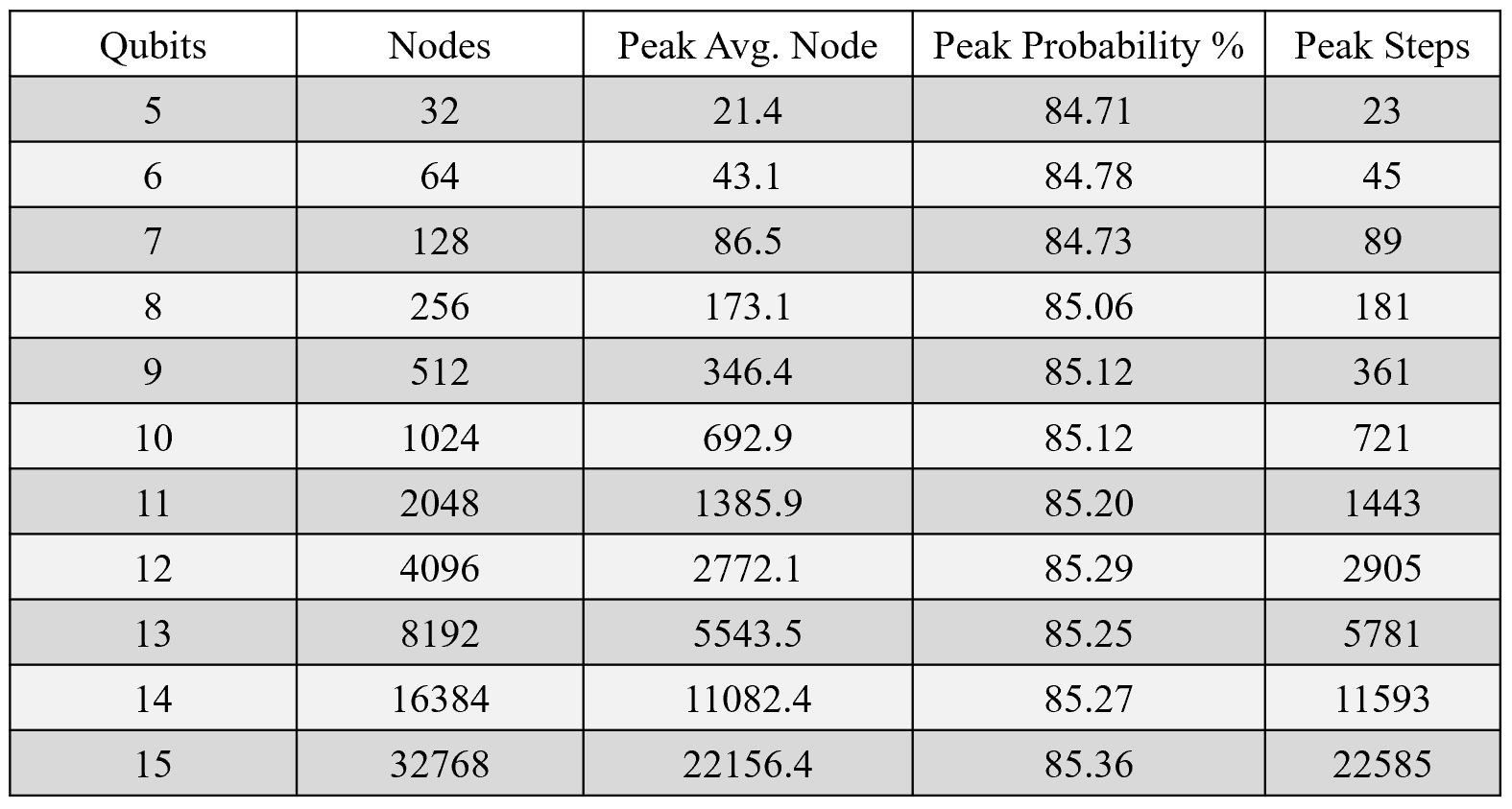}
		\caption{A table of values showing statistics of one dimensional Coin Boundary walks as node qubits increase.  `Peak Avg. Node' refers to the largest value found according to equation \ref{Eqn: Node Prob}. `Peak Probability \%' documents the total probability in the system accumulated on nodes $\geq 2^{N-1}$ (the right half of the walk) at the moment of the peak. `Peak Steps' is the number of steps each quantum walk needs in order to reach the peak average node probability, which is directly proportional to the size of the walk. }
		\label{Fig: CB Table}
	\end{figure}

	A closer look at the three probability distributions in figure \ref{Fig: CB Prob Plot} reveals how the effect of the Coin Boundaries propagate from both sides with a speed of 1 node per step.  Thus, nodes in the center do not feel the effect of the Coin Boundaries until much later, shown by plots for steps 22 and 45.  Because of this propagation speed, the `peak' of the walk, or the moment when the average node probability is largest, is always around the moment when the effects from both boundaries collide at the centermost nodes.  When this happens, the two effects begin to cancel one another, and eventually leads to the majority of probability concentrating on the left half of the walk.  Through simulations of these bounded walks, we were able to determine that the entire process is cyclic, whereby all of the probability in the system continually shifts back and forth between the two boundaries, without ever decaying in peak amplitudes.  Table \ref{Fig: CB Table} below provides some numerical context to exactly how these walks are behaving, most notably their peak properties.
	
	An important notable from table \ref{Fig: CB Table} is the `Peak Steps' column, which shows the quarter-cycle period for each walk size.  In particular, going from an $N$ to $N+1$ sized quantum walk causes the overall period to almost exactly double.  This is in agreement with the observation shown by the step 22 plot of figure \ref{Fig: CB Prob Plot}, where doubling the size of the walk should in principle double the time it takes for the effects from each boundary to collide in the center.  Also, the 'Peak Probability \%' column of the figure, which records how much probability is concentrated in the right half of the walk, is an interesting phenomenon for algorithmic purposes.  Knowing the where, and when over 85\% of the total probability in a system is concentrated could be a potentially powerful tool for solving the right kind of problem \cite{koch_scatter_1,koch_scatter_2}.
	
	Looking at the top left plot of figure \ref{Fig: CB Prob Plot} once more, the overall shape of the average node probability plot appears sinusoidal in nature, especially due to its oscillatory structure with predictable periods.  However, upon further investigation, we found that the nature of the probability cycle is in fact much closer to quadratic.  More specifically, the walk propagates between two seemingly quadratic curves, which can be seen in the jaggedness of figure \ref{Fig: CB Prob Plot}.  As evidence of this underlying quadratic structure, figure \ref{Fig: Best Fits} below shows best-fit approximations to the upper and lower curves, which minimize the quantity $\sigma$ provided in equation \ref{Eqn: Sigma} (Y$_i$ are the data points generated from the walk and Y$_i'$ are the best-fit).

	\begin{eqnarray}	
		\sigma \hspace{.05cm} \equiv \hspace{.05cm} \sqrt{ \frac{ \sum_i ( \textrm{Y}_i - \textrm{Y}_i' )^2 }{N} }
		\label{Eqn: Sigma}
	\end{eqnarray} 
	
	\begin{figure}[h]                     
		\centering
		\includegraphics[scale=0.5]{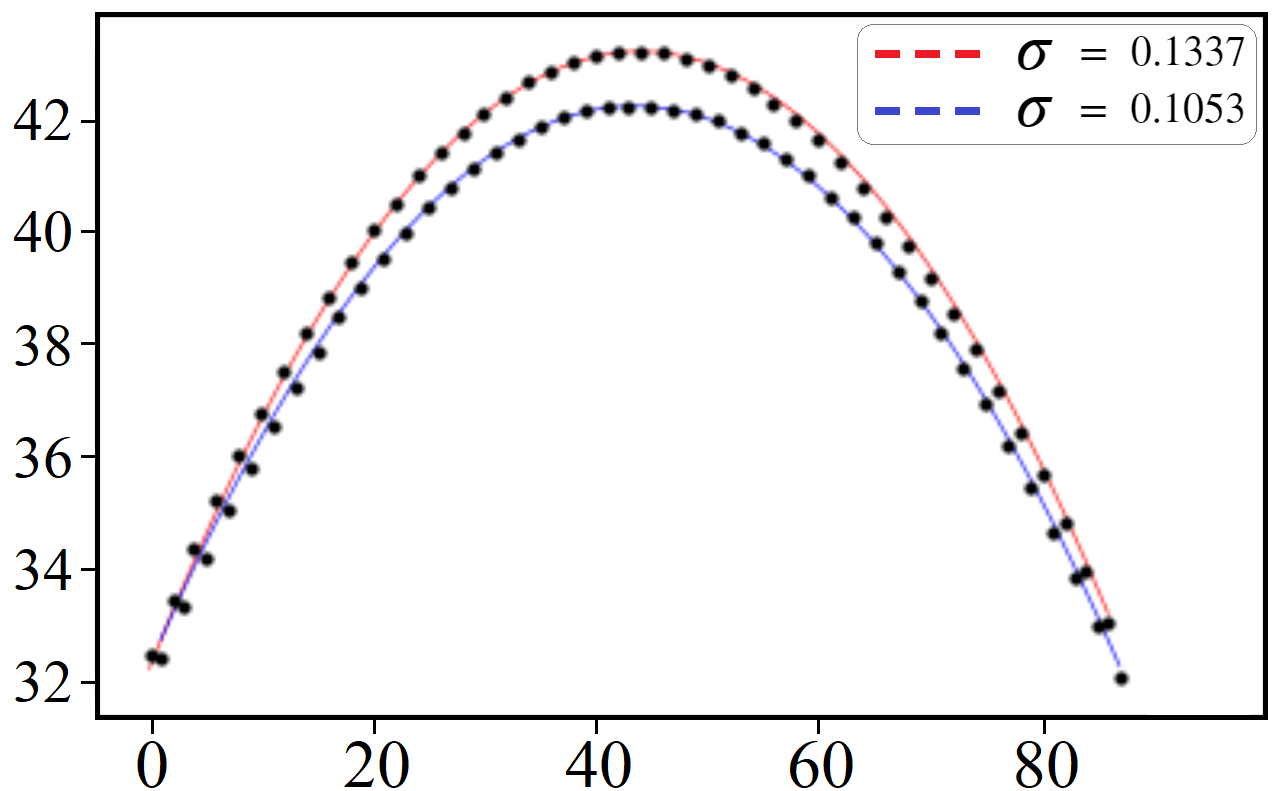}
		\caption{ (dashed lines) Parabolic best-fits to average node probability data for the case of a 6-qubit bounded walk. (black points) Data points generated using a classical simulator (Qiskit) representing a noiseless walk. (top right) $\sigma$ values for each best fit, according to equation \ref{Eqn: Sigma}}
		\label{Fig: Best Fits}
	\end{figure}
	
	For reference, sinusoidal best-fits were also performed using the same data points for figure \ref{Fig: Best Fits}, as well as other walk sizes, yielding $\sigma$ values around 0.2 - 0.3 higher.  As for the underlying nature behind why these plots are quadratic and not sinusoidal, we leave this question as an open topic for future research.  We hypothesize that the quadratic nature is tied to Hadamard Coin, and the way in which interference effects are happening at both ends of the walk.  Because the propagation speed of the boundary effects is always 1 node/step regardless of the coin, this leads us to believe that different coins could potentially lead to similar peaking effects, but with different analytical structures (possibly other polynomial forms).  However, further investigation is needed.
	
	\section{Constructing Higher Dimensional Walks}
	
	In light of the various circuit design techniques demonstrated thus far, here we present findings on the scalability of these walks into higher dimensions.  In this study, we explore a strategy based on the way in which QADD operations of different sizes (see figure \ref{Fig: QADD3}) can be used to achieve different arithmetic operations of $\pm 2^N$.  By separating the total number of node qubits into subgroups based on desired dimensional length, shifts in orthogonal directions can be achieved through QFT and QADD operations on these subgroups of qubits.  Figure \ref{Fig: 4x4} below shows a two dimensional example, demonstrating a 4x4 grid and its corresponding node numbering.
	
	\begin{figure}[h]                     
		\centering
		\includegraphics[scale=.25]{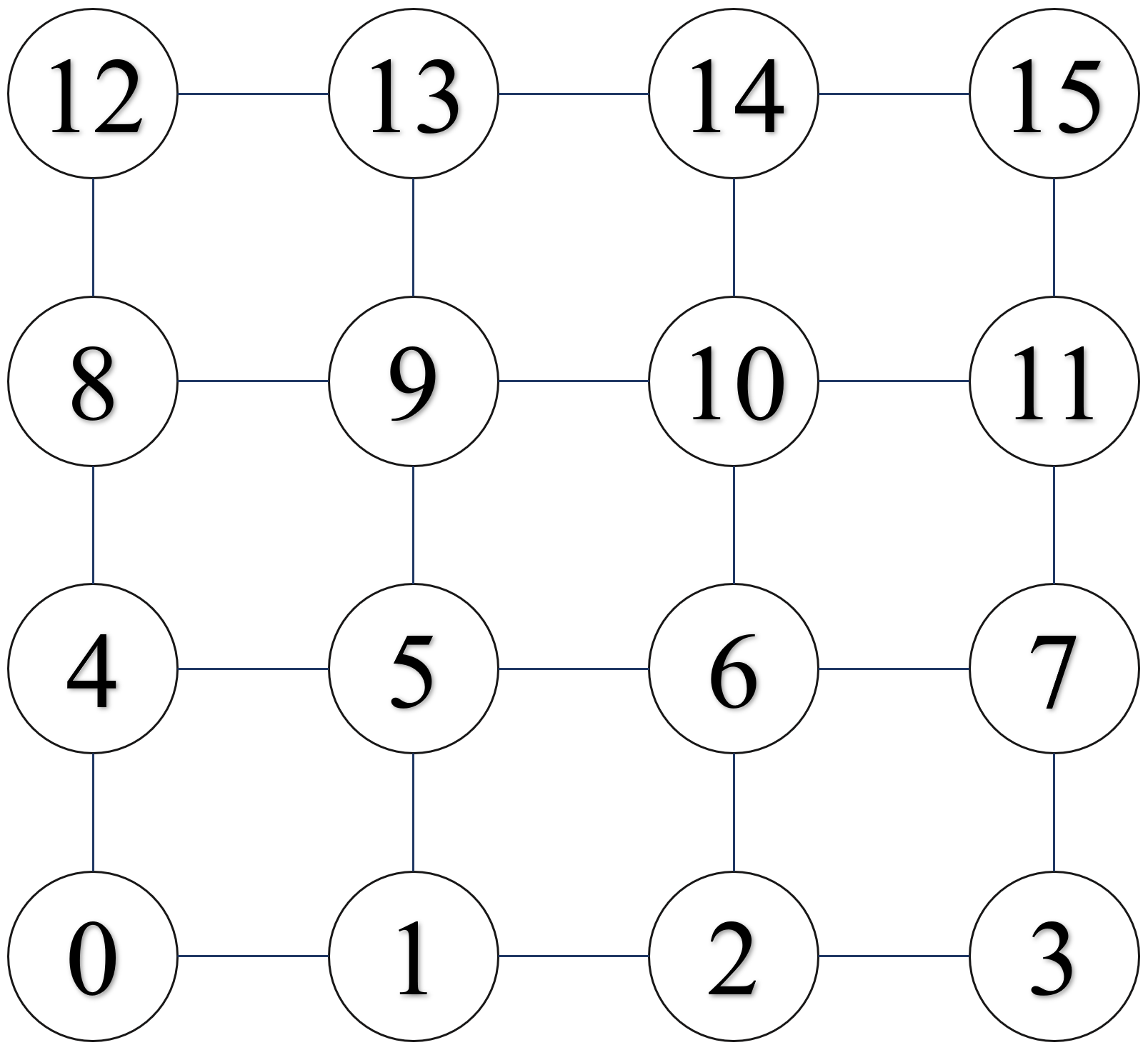}
		\caption{A 2-dimensional grid containing a total of 16 nodes, created from 4 qubits.}
		\label{Fig: 4x4}
	\end{figure}
	
	In figure \ref{Fig: 4x4}, note that all shifts in the horizontal direction correspond to modulo arithmetic operations of $\pm1$, while vertical shifts are achieved through $\pm4$ (period boundaries in both directions).  Given a particular node state $|N\rangle$, represented by the binary number state $|\textrm{Q}_1 \textrm{Q}_2 \textrm{Q}_3 \textrm{Q}_4 \rangle $ (rightmost least significant bit), one can achieve horizontal shifts by applying QADD$^{\pm}$ to the subsystem $|\textrm{Q}_3 \textrm{Q}_4 \rangle$, and similarly vertical shifts on the subsystem $|\textrm{Q}_1 \textrm{Q}_2 \rangle$.  This is because a $\pm1$ change on the state $|\textrm{Q}_1 \textrm{Q}_2 \rangle$ results in a $\pm4$ change for the total binary state $|\textrm{Q}_1 \textrm{Q}_2 \textrm{Q}_3 \textrm{Q}_4 \rangle $.  And in general, this strategy can be applied to create any N-dimensional shift by adding on additional subgroups of qubits.
	
	\begin{figure}[h]                     
		\centering
		\includegraphics[scale=.4]{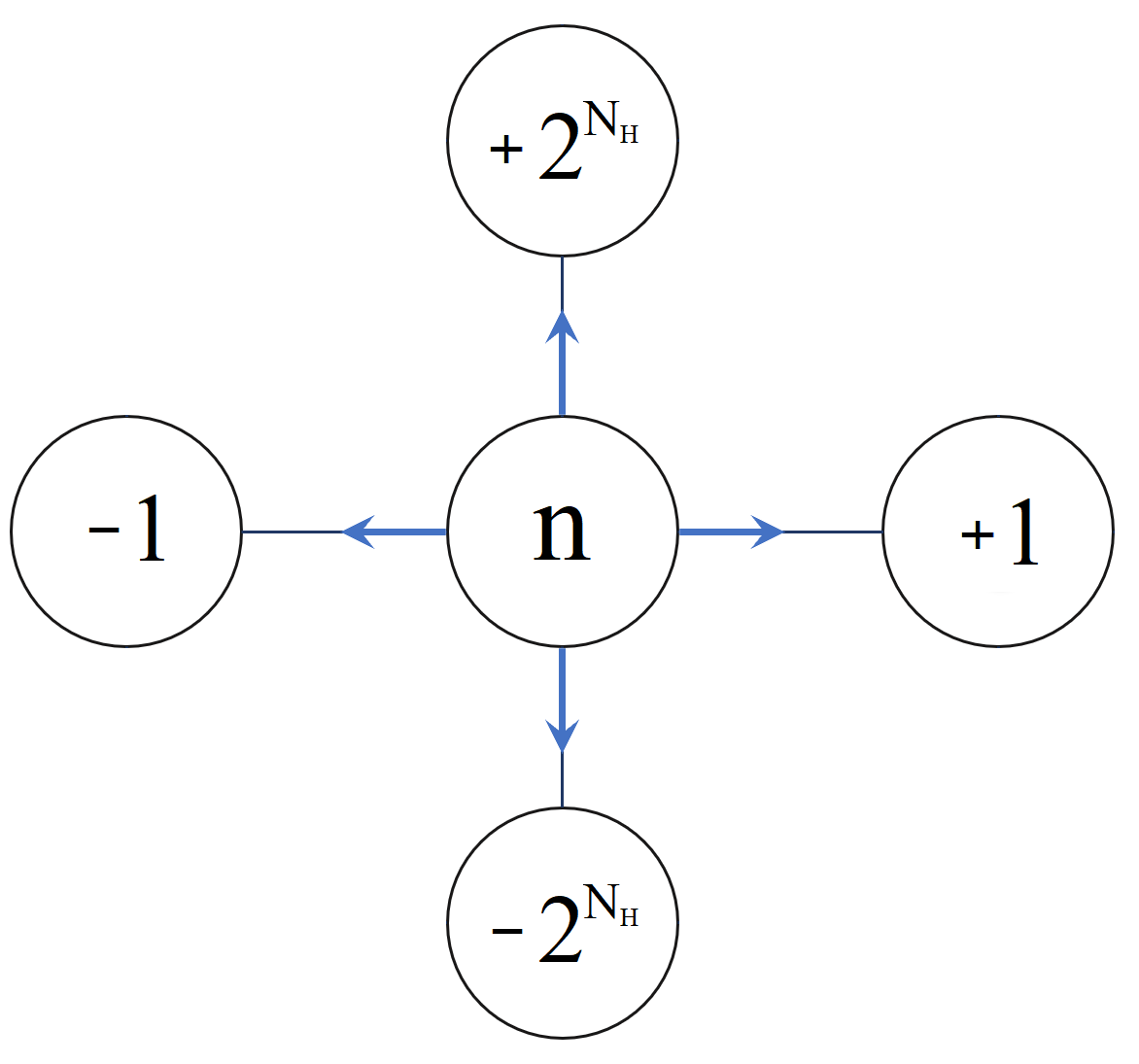}
		\caption{Modulo arithmetic operations corresponding to horizontal and vertical shifts for a two dimensional quantum walk, where N$_{\textrm{H}}$ is the number of horizontal qubits in the system. }
		\label{Fig: TDOS}
	\end{figure}
	
	\subsection{2D Shift Operator}
	
	When creating higher dimensional walks through separation of qubits into subsystems, at least one additional coin qubit is needed per dimension in order to create orthogonal coin states for shifting.  In general, an N-dimensional walk requires at least N coin qubits to create $2^N$ orthogonal coin states.  And how one defines shifts based on these $2^N$ coin states is a degree of freedom, just like the $|0\rangle$ and $|1\rangle$ states in one dimension.
	
	\begin{eqnarray}                 
		\textrm{U}_S  && =  \sum_{n} | n_{\uparrow}  \rangle \langle n|  \otimes | \textrm{C}_{\uparrow} \rangle \langle \textrm{C}_{\uparrow} |  +  | n_{\downarrow}  \rangle \langle n|  \otimes | \textrm{C}_{\downarrow} \rangle \langle \textrm{C}_{\downarrow} | \nonumber \\
		+&& | n_{\rightarrow}  \rangle \langle n|  \otimes | \textrm{C}_{\rightarrow} \rangle \langle \textrm{C}_{\rightarrow} |  +  | n_{\leftarrow}  \rangle \langle n|  \otimes | \textrm{C}_{\leftarrow} \rangle \langle \textrm{C}_{\leftarrow} |
		\label{Eqn: 2D Shift Operator}
	\end{eqnarray} 
	
	In equation \ref{Eqn: 2D Shift Operator}, the four coin states $| \textrm{C}_i \rangle $ can be chosen as any set of four orthogonal 2-qubit states.  Additionally, the four numerical shifting directions $| n_i \rangle$ do not necessarily have to represent the four directions depicted by the grid lines in figure \ref{Fig: 4x4}, so long as every state in the system can reach all other states through repetitions of mixing and U$_S$ (for example, strictly diagonal shifting cannot reach all states on a 2D grid with periodic boundaries).  Figure \ref{Fig: 2D Dep Coin} below illustrates one such example.  Each of the four directions is achieved through QADD$^{\pm}$ operations acting on the vertical / horizontal qubit subsystems, labeled N$_\textrm{V}$ and N$_\textrm{H}$ respectively.  Just like the one dimensional case, X gates are used to isolate each of the four coin basis states $|00\rangle$, $|01\rangle$, $|10\rangle$, and $|11\rangle$ for use as controls to the QADD$^{\pm}$ operations.
	
	\begin{figure}[h]                     
		\centering
		\includegraphics[scale=.3]{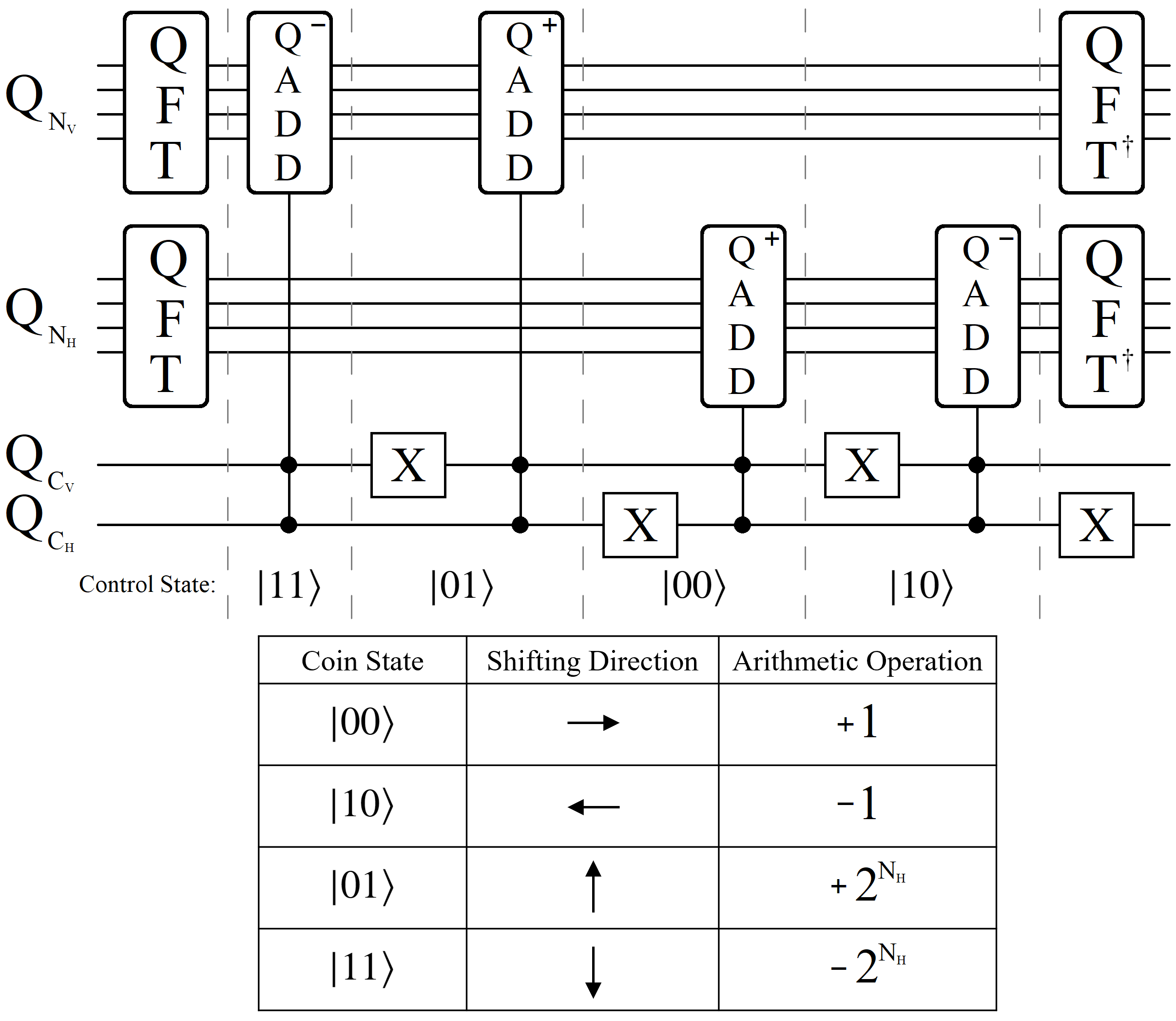}
		\caption{ (top) Quantum circuit for a two dimensional Shift Operator which results in shifts corresponding to figure \ref{Fig: TDOS}.  In order for each of the 2-qubit coin basis states to produce unique shifting directions, both of the coin qubits are required to serve as controls for the various QADD$^{\pm}$ operations on the vertical and horizontal subsystems. (bottom) Table summarizing the four shift operations, their control coin states, and corresponding modulo arthimetic operations.}
		\label{Fig: 2D Dep Coin}
	\end{figure}
	
	The circuit shown in figure \ref{Fig: 2D Dep Coin} represents the complete Shift Operator for one step of a two dimensional quantum walk.  It follows from the same logic as the one dimensional case depicted in figure \ref{Fig: Core Walk}, but comes at the cost of two important circuit requirements: 1) Double the gate count and total circuit length.  2) Both of the coin qubits must serve as controls for each of the four QADD$^{\pm}$ operations.  The first of these two requirements is to be expected of higher dimensional walks, as each extra dimension to the walk should bring with it additional circuitry.  However, the second point is the more problematic of the two, as higher order control operations require extra ancilla qubits, further adding to the issue of connectivity outlined in section II.  One advantage to choosing the four shifting coin states in the manner shown in figure \ref{Fig: 2D Dep Coin} is that each pair of opposite directional states only differ by the first coin qubit (for example, $|00\rangle$ and $|10\rangle$ for right and left).  Consequently, this convention will make enforcing boundaries in 2D more gate efficient, which we discuss further in the next subsection.

	\subsection{2D Coin Boundaries}
	
	Following from the methodology outlined in figures \ref{Fig: Coin Boundary Walk} and \ref{Fig: CB Example}, here we present similar results on how to enforce various  boundary conditions in two dimensions.  But first, returning briefly to the motivation behind boundaries, figure \ref{Fig: Grid Graph} below illustrates how one could theoretically create arbitrarily complex graph structures using the boundary techniques discussed in this section.  A primary focus of this study is to provide fundamental tools from which desirable quantum walk geometries can be constructed, whether they are currently NISQ-friendly or not.  The hope is that through continued research, the promise of quantum walks can transition from mathematical constructs to physically implementable circuits with standard gate sets.
	
	\begin{figure}[h]                     
		\centering
		\includegraphics[scale=.3]{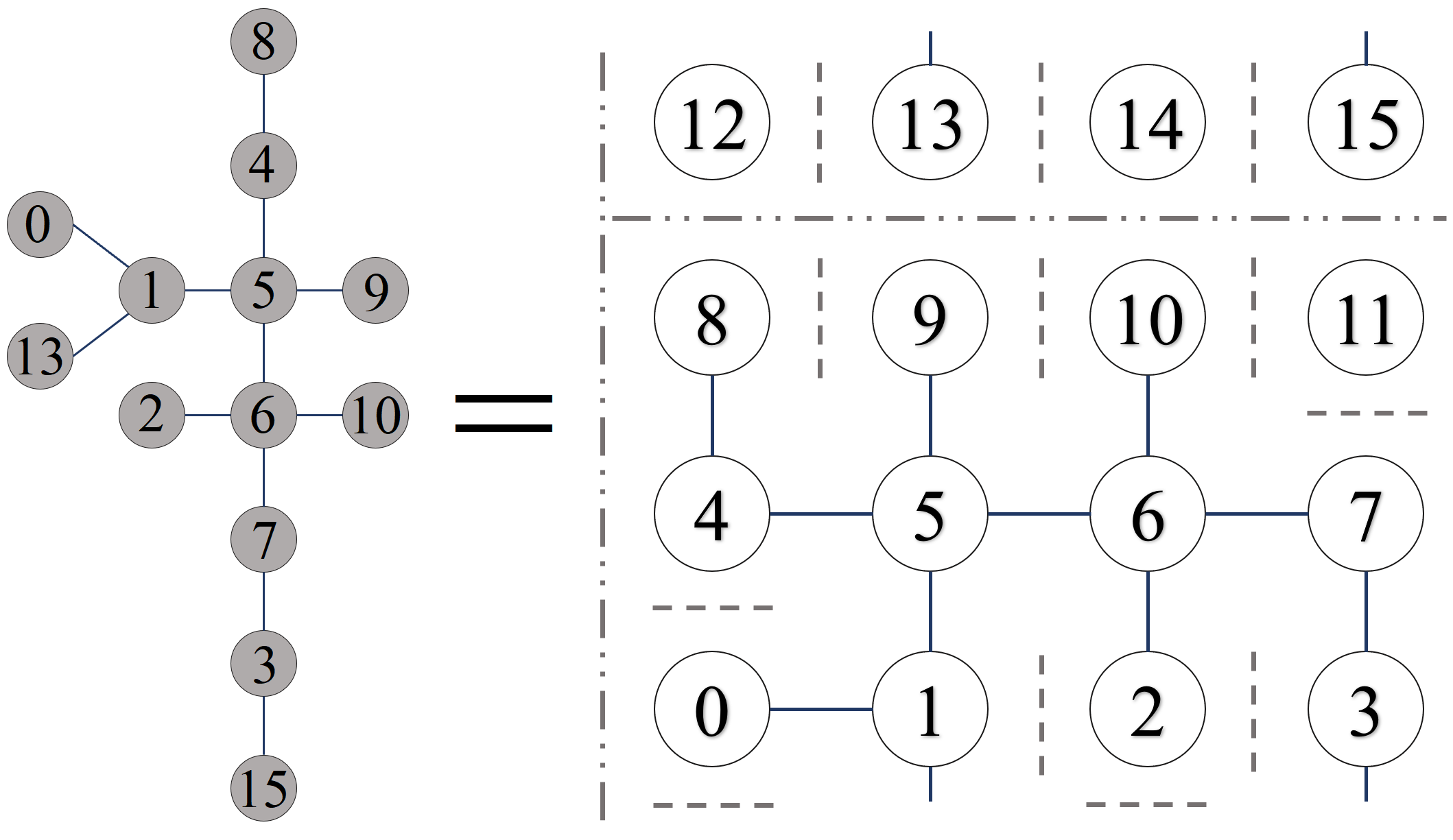}
		\caption{ Example of how one might create a quantum walk on a desirable graph structure using single state boundaries (single dash) and global boundaries (alternating dash). }
		\label{Fig: Grid Graph}
	\end{figure}
	
\begin{figure}[h]                     
		\centering
		\includegraphics[scale=.35]{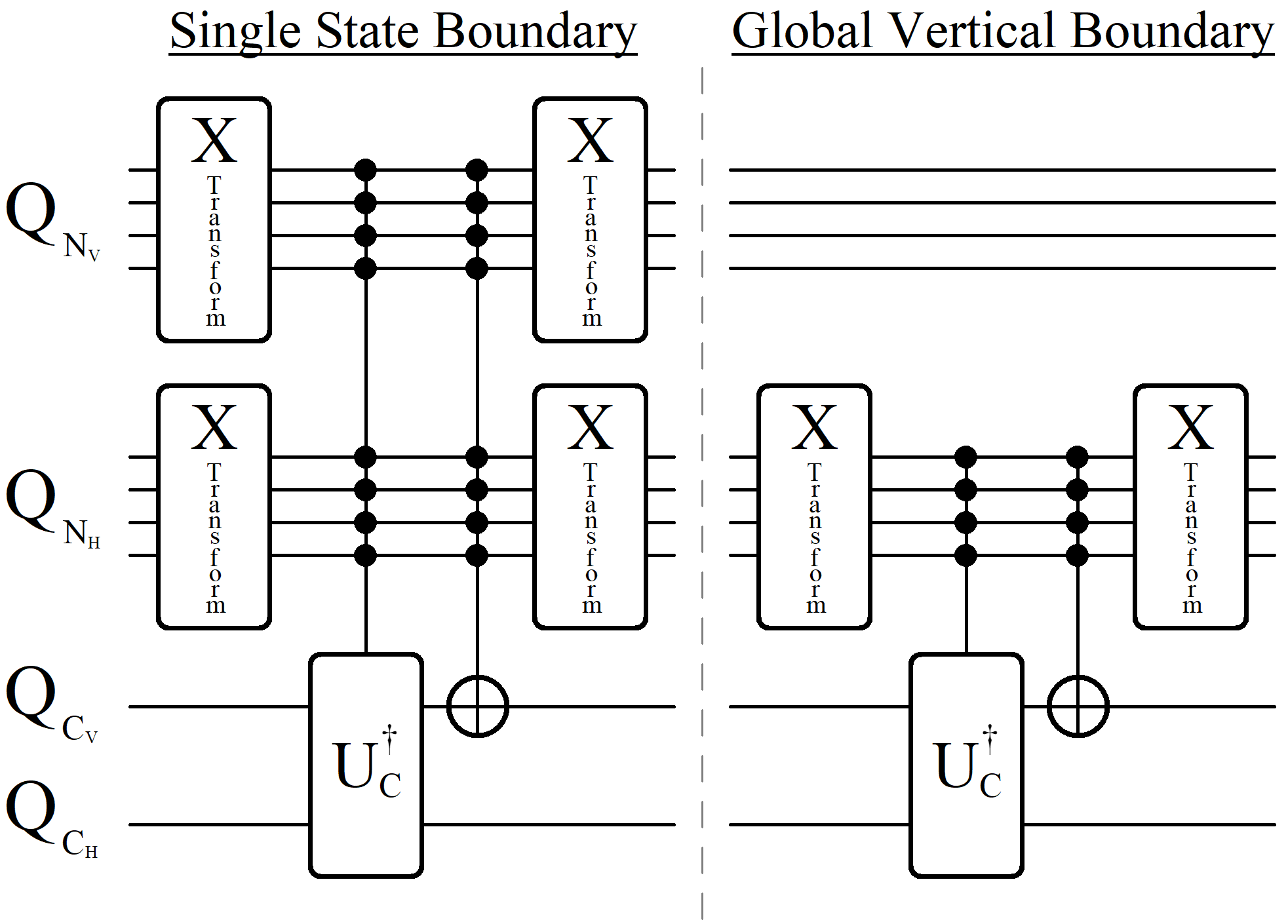}
		\caption{ Quantum circuits for implementing a single state boundary (left) or global boundary (right) for a two dimensional quantum walk.}
		\label{Fig: 2D CB Circ}
	\end{figure}

	To begin, the circuitry shown in figure \ref{Fig: Coin Boundary Walk} can be directly applied to either vertical or horizontal  qubit subsystem to create a global boundary, preventing all vertical or horizontal shifts across node states which share the same binary state component.  Using figure \ref{Fig: Grid Graph} as an example, applying a one-dimensional coin boundary between nodes 0 and 3 will also result in boundaries between nodes 4 \& 7, 8 \& 11, and 12 \& 15 as well (see figure \ref{Fig: 4x4}).  This is because all four pairs of nodes share the same horizontal qubit component: states $|01\rangle$ and $|10\rangle$, only differing in their vertical qubit states.  Thus, a control chain only stemming from only the horizontal qubit subsystem, like shown in figure \ref{Fig: 2D CB Circ}, results in vertical boundaries for numerous states.
	
	While global boundaries can be thought of as an efficient way for achieving numerous boundaries with minimal gates and connectivity (only requiring operations on one subsystem of qubits), more complex graph structures require localized single state boundaries.  However, achieving these boundaries comes at the cost of now requiring control chains stemming from both the vertical and horizontal qubit subsystems, acting on both coin qubits, in order to apply U$^{\dagger}_C$. Figure \ref{Fig: 2D CB Circ} below shows an example of this for the general case of a 2-qubit U$_C$ operator.  
	
	The mathematical strategy behind single state boundaries in two dimensions is largely the same as the one dimensional case, whereby the goal is to undo the mixing effect of U$_C$ at the location of the boundary (on both sides), followed by a second transformation to ensure the forbidden coin state(s) is no longer present before shifting.  In two dimensions however, a single node can be subject to up to three different boundaries simultaneously, leading to 14 unique configurations.  However, we demonstrate that by choosing ideal coin states for shifting, such as figure \ref{Fig: 2D Dep Coin}, all 14 geometries can be handled by the same X transformation for enforcing reflective boundaries.  Shown below in figure \ref{Fig: 2D Boundary Configs} are three unique geometric cases, their anticipated incoming states, and the final outgoing states as a result of an X gate on only the first coin qubit.
	
	\begin{figure}[h]                     
		\centering
		\includegraphics[scale=.32]{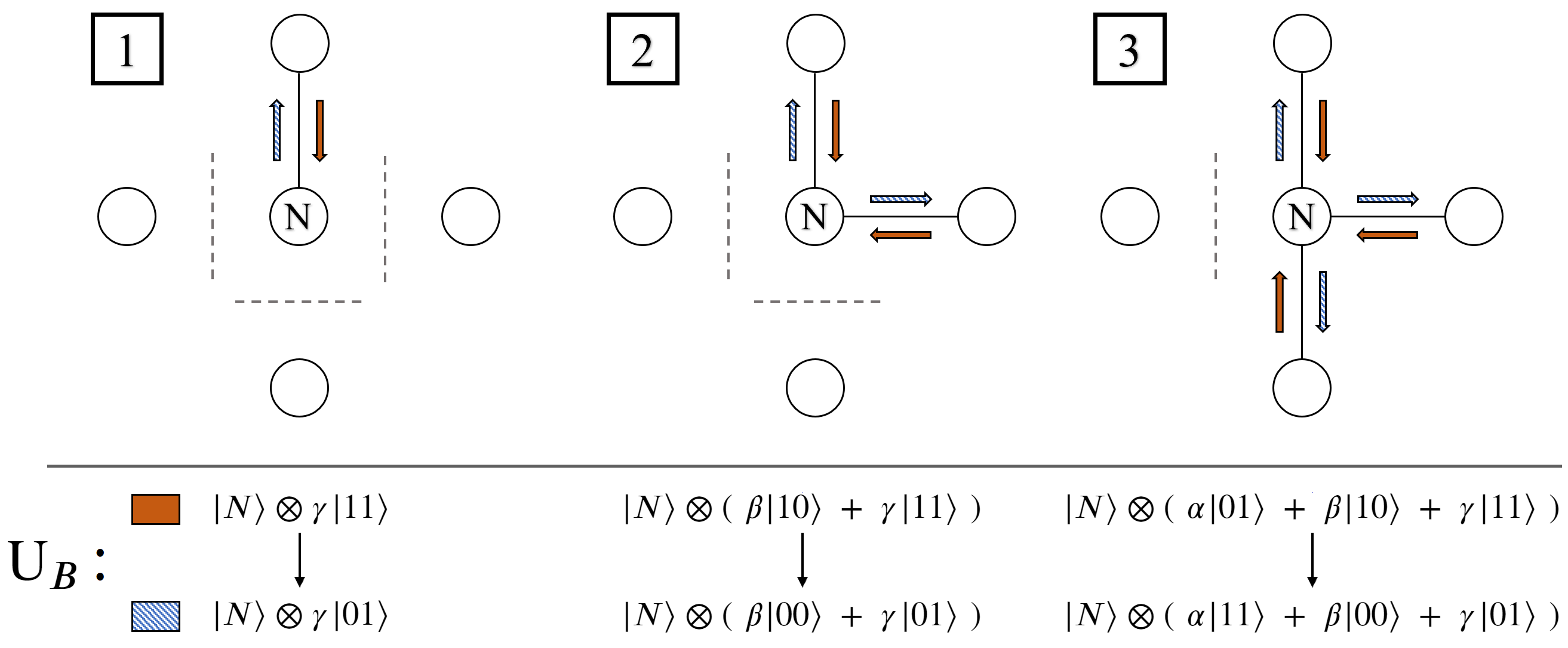}
		\caption{Examples of the possible 14 configurations which can arise from single state boundaries, whereby a node is surrounded by three (left), two (middle), or one boundary (right).  By choosing to implement the four directional shifts according to figure \ref{Fig: 2D Dep Coin}, all 14 boundary configurations can be handled by flipping the state on only one coin qubit.}
		\label{Fig: 2D Boundary Configs}
	\end{figure}
	
	In all of the examples shown in figure \ref{Fig: 2D Boundary Configs}, applying an X gate to the first coin qubit results in new coin states which represent shifts in the opposite direction ( up \& down, and left \& right ).  If one chooses to assign different directional shifts to the coin basis states, the consequence may require X operations on both coin qubits.  Additionally, analogous to the phase factor available to one dimensional boundaries (equation \ref{Eqn: 1D Boundary U}), enforcing boundaries in two dimensions comes with new degrees of freedom.  Using the case of a single boundary as an example, equation \ref{Eqn: 2D Boundary U} shows the general requirement for the boundary condition:
	
	\begin{eqnarray}                 
		|\Psi\rangle &=&  \hspace{.02cm} |N\rangle \otimes \big{(} \hspace{.06cm} \alpha |01\rangle + \beta |10\rangle + \gamma |11\rangle     \hspace{.06cm} \big{)}  \nonumber \\
		\textrm{U}_B &\rightarrow&  \hspace{.02cm} |N\rangle \otimes \big{(} \hspace{.06cm} \alpha' |01\rangle + \beta' |00\rangle + \gamma' |11\rangle     \hspace{.06cm} \big{)} 
		\label{Eqn: 2D Boundary U}
	\end{eqnarray} 
	
	The states shown in equation \ref{Eqn: 2D Boundary U} correspond to the rightmost case in figure \ref{Fig: 2D Boundary Configs}, whereby a single boundary is placed to block leftward shifts.  In general, the only requirement of this boundary condition is that the coin state $|10\rangle$ be removed from the final superposition state, which earlier was achieved via an X gate on the first coin qubit, transforming it to $|00\rangle$.  However, how the amplitudes between the states $|01\rangle$ and $|11\rangle$ propagate is a degree of freedom.  If desirable, these amplitudes can be made to entirely reflect as well, fully transmit through to one another, or some degree of partial reflection / transmission along with new phases.  Invoking these various conditions requires additional control operations stemming from the coin qubits themselves, but so long as the net effect is describable by equation \ref{Eqn: 2D Boundary U}, any U$_B$ operation can be used.
	
	The final point mentioned above may be subtle, but it encompasses the general strategy behind using coin boundaries.  In equation \ref{Eqn: 2D Boundary U}, note that the amplitude $\beta$ on the incoming state $|10\rangle$ is not required to be exclusively transformed into the state $|00\rangle$, which can have a new amplitude $\beta'$.  The example shown in figure \ref{Fig: 2D Boundary Configs} represents the most efficient U$_B$, the case of full reflection via a single control-X gate, but in principle more complex unitary transformations can allow for mixing at node locations with adjacent boundaries (for example unitary rotation operations).  
	
	\subsection{IBM Implementation}
	
\begin{figure}[h]                     
		\centering
		\includegraphics[scale=.3]{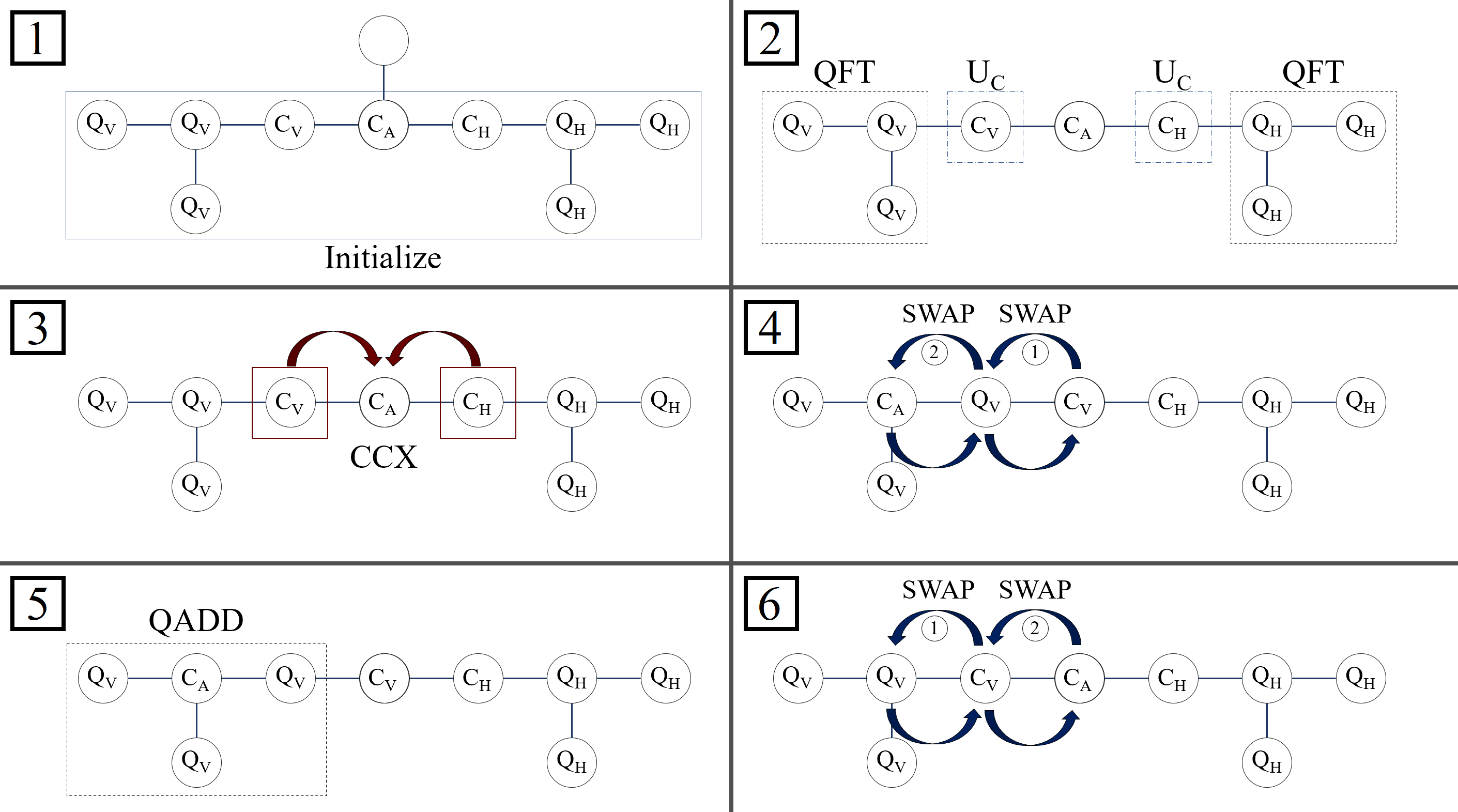}
		\caption{Steps for implementing a two dimensional quantum walk on IBM's heavy hexagonal geometry, using 3 vertical (Q$_V$), 3 horizontal (Q$_H$), 2 coin (C$_V$ \& C$_H$), and 1 ancilla (C$_A$) qubit.  Shown in steps 3-6 are the necessary operations for achieving a downward vertical shift on all states tensored to the coin state $|11\rangle$. Similarly, the remaining three shifts can be handled by repeating steps 3-6 with the addition of X Transformations on the coin qubits according to figure \ref{Fig: 2D Dep Coin}.}
		\label{Fig: HH 2D}
	\end{figure}

	As the final topic for this study, here we present a few comments on how one might go about implementing higher dimensional walks onto IBM's heavy hexagonal architecture.  Analogous to the circuit in figure \ref{Fig: HH QFT}, the main strategy is to optimize qubit locations / connectivity throughout the circuit using SWAP gates. Figure \ref{Fig: HH 2D} below outlines steps for a 6-qubit walk, composed of 3 vertical and horizontal qubits, creating an 8x8 node graph structure.  Each of the 3-qubit vertical and horizontal subsystems is located at different 4-qubit junctions on the chip \cite{ibmq_tor}, with three intermediate qubits separating the central qubits from each junction.  These intermediate qubits serve as coins for the walk, one for each subsystem, with one additional ancilla qubit (C$_A$) necessary for shifting.
	
	As a consequence of the two dimensional Shift Operator outlined in figure \ref{Fig: 2D Dep Coin}, requiring both coin qubits to serve as control means that an additional ancilla qubit is needed to absorb their joint CCX (Toffoli gate) operation, shown in step 3.  X gates can be used to enforce the four 2-qubit coin basis states as control, which in turn create the four orthogonal shifts via QADD$^{\pm}$ operations stemming from the ancilla qubit.  However, moving the state of the ancilla qubit C$_A$ around the geometry is the bottleneck of the circuit, requiring four SWAP gates per directional shift (steps 3 \& 5).

	The steps outlined in figure \ref{Fig: HH 2D} represent the general strategy for implementing higher dimensional walks, but there is potential for optimizations which could lead to higher fidelities.  The most notable being the extra unused qubit shown in step 1, connected to qubit C$_A$.  Without using this extra qubit, the most costly bottleneck of the circuit is moving the state of C$_A$ around the chip via SWAP gates, which must travel between the two qubit subsystems. This means that only one of the node qubit subsystems can receive a QADD$^{\pm}$ operation at a time, leaving the majority of qubits in the system idle, which in turn makes the algorithm very prone to decoherence effects \cite{T1_1,T2_1}.  However, below in figure \ref{Fig: HH 2D Improved} we propose an optimization technique which makes use of the unused ancilla qubit to reduce circuit length. 
	
	\begin{figure}[h]                     
		\centering
		\includegraphics[scale=.3]{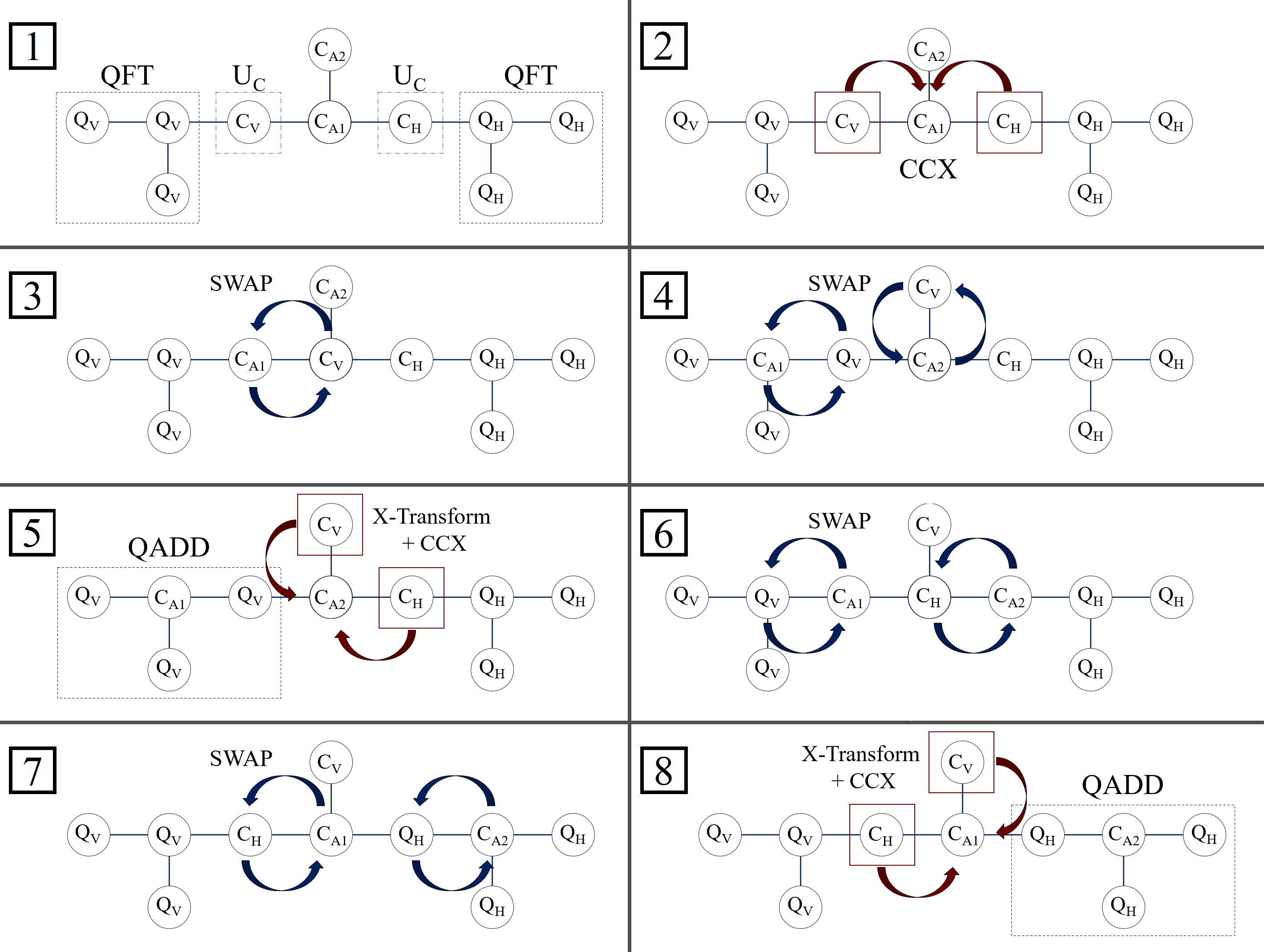}
		\caption{Steps for implementing a two dimensional quantum walk on IBM's heavy hexagonal geometry, using two ancilla qubits to minimize circuit length.  The state of the two ancilla qubits, C$_{A1}$ and C$_{A2}$, alternate between sharing connectivity with the coin qubits and the node qubit subsystems via SWAP gates, minimizing idle time as compared to the approach laid out in figure \ref{Fig: HH 2D}.}
		\label{Fig: HH 2D Improved}
	\end{figure}
	
	The circuit design outlined in figure \ref{Fig: HH 2D Improved} is unique in that it utilizes an extra ancilla qubit in order to try and increase fidelity, which is typically the opposite approach to circuit optimization.  But steps 5 - 8 in the figure illustrate the strength of the technique, whereby one ancilla qubit is delivering the QADD$^{\pm}$ operation to one subsystem, while the other ancilla qubit is simultaneously being prepared for the next QADD$^{\pm}$.  This process is then repeated until all four shift operations are performed, effectively cutting idle qubit time down by half.
	
	Despite significantly reducing overall circuit length, it isn't immediately clear if the optimization proposed in figure \ref{Fig: HH 2D Improved} $\textit{does}$ result in better fidelities. Unfortunately, at the time of this study, performing both of the quantum circuits corresponding to figures \ref{Fig: HH 2D} and \ref{Fig: HH 2D Improved} are just slightly out of reach for IBM's quantum volume 32 Toronto chip.  Preliminary tests of both circuits yielded fidelities below 10\%, making a comparison between them difficult.  However, as future generations of better qubits are made available, we plan to revisit these circuit techniques for a more detailed investigation of their fidelities.

	\section{Conclusion}
	
	The techniques and results demonstrated throughout this study highlight the complexity of implementing non-trivial quantum walks.  Overall, we found that a strong case can be made for the Quantum Adder Shift Operator as a candidate for quantum walks on superconducting qubits.  The use of phase space to achieve arithmetic operations on large superposition states is the technique's primary strength, but it is also it's most costly aspect.  The QFT and QFT$^{\dagger}$ requirements create the largest demand for connectivity between qubits, add additional gates / circuit depth, and perhaps most importantly, create a division within the algorithm which bars the Shift Operator's action from the rest of the quantum circuit.  However, many quantum algorithms find themselves in this same boat  (and similarly classically with FFT based algorithms in parallel computing), bottleneck by QFT, so it is reassuring to know that there are research efforts continually investigating better / more efficient ways for QFT implementations \cite{qft_better,qft_better2,qft_better3}.  As these advancements continue to be discovered and refined, they in turn boost the viability of this particular Shift Operator, and quantum walks overall.
	
	It is our conclusion that the true strength of the Quantum Adder Shift Operator is in its simplicity, and scalability into higher dimensions.  The use of qubit subsystems for representing nodes / shifts in orthogonal directions is a positive trait for NISQ computers with limited connectivity.  In sections III and V we demonstrated this principle for circuit's tailored to IBM's heavy hexagonal connectivity, using SWAP gates to maximize connectivity potential for qubits during the QFT and Shift operations.  Speculating forward now, in order for this style of Shift Operator to obtain any true speedups, the gate/ circuit depth cost of QFT (or multiple QFT's) needs to be outweighed by the $2^N$ node scaling power one gets from using $N$ qubits (versus classical O($N$) storing of node information).  To date, this requires much larger scale quantum walks than currently achievable on NISQ computers with limited connectivity.
	
	\subsection{Future Work}
	
	As is the nature of any quantum algorithm pursuit, there is always more work to be done until a quantum speedup is obtained, and quantum walks are no different.  However, the path to a quantum walk based speedup requires both experimental and algorithmic advancements. Technological breakthroughs in qubit technologies allow for new possibilities of implementation, while algorithm designs must strive to fully utilize every advantage a technology has to offer.  
	
	Based on the results of this study, we identify two avenues of future research to better understand the potential of these superconducting qubit quantum walks.  Firstly, more exploration is needed into the different kinds of wavefunctions / probability distributions which can arrive from combining different coins, geometries, boundaries, etc.  Discoveries of advantageous new probability distributions can spark new algorithmic ideas, and vice versa.  Secondly, achieving Shift Operators through means different from this study (Quantum Adder Algorithm) is an important unanswered question for better circuit designs.  Having a non-QFT Shift Operator could be a huge boost in circuit efficiency, and potentially unlock new techniques for boundaries and other graph complexities.
	
	\section*{Acknowledgments}
	
	We gratefully acknowledge support from the National Research Council Associateship Program.  We would also like to thank the IBM Quantum Experience team and all of their support.  Any opinions, findings, conclusions or recommendations expressed in this material are those of the author(s) and do not necessarily reflect the views of AFRL.
	
	\section*{Data \& Code Availability}
	The data and code files that support the findings of this study are available from the corresponding author upon reasonable request.

	\pagebreak
	
	
	\pagebreak
	
	\appendix	
	\section{Appendix A}
	
	\begin{figure}[h]                     
		\centering
		\includegraphics[scale=.4]{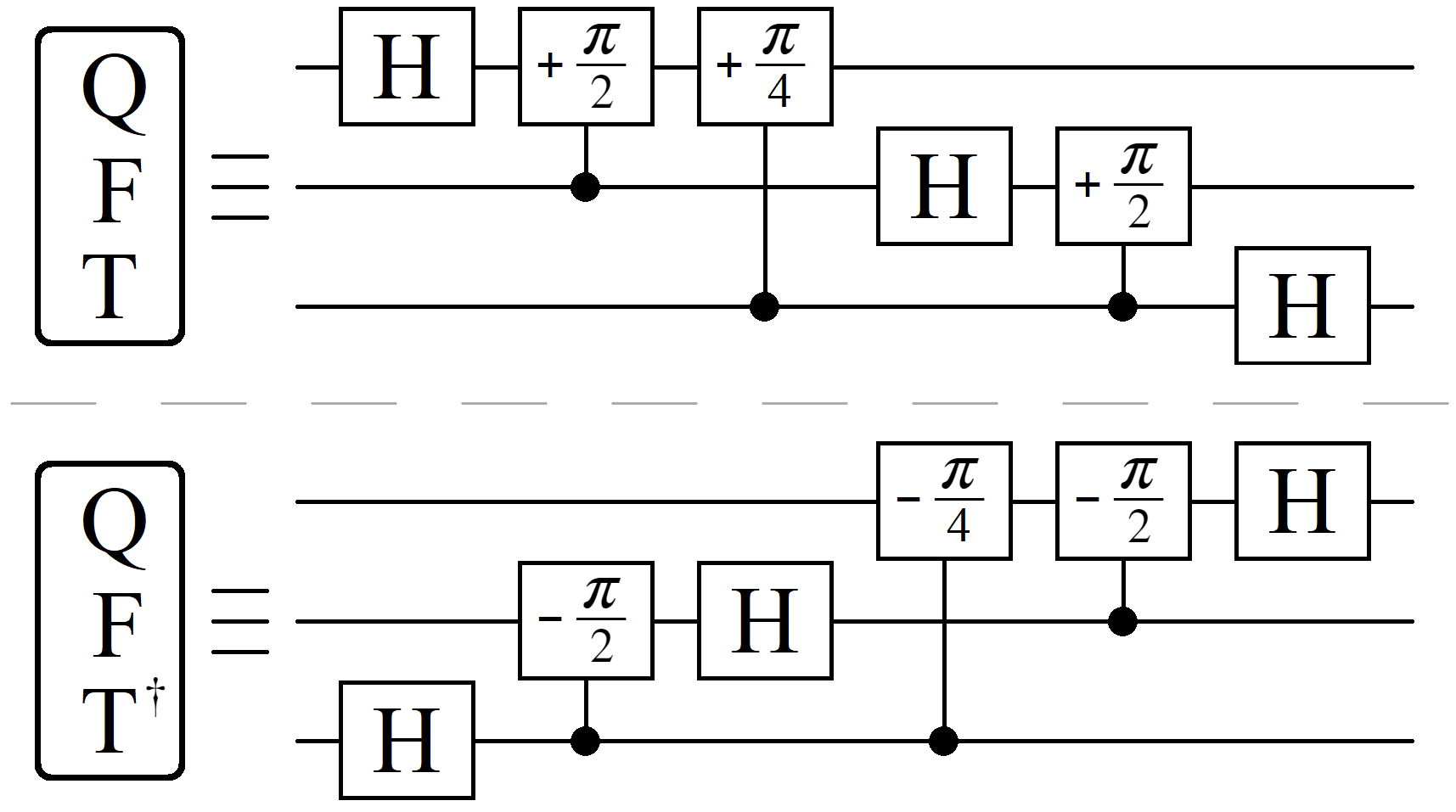}
		\caption{Three Qubit QFT and QFT$^{\dagger}$ quantum circuits}
		\label{Apx: qftc}
	\end{figure}
	
	\begin{figure}[h]                     
		\centering
		\includegraphics[scale=.25]{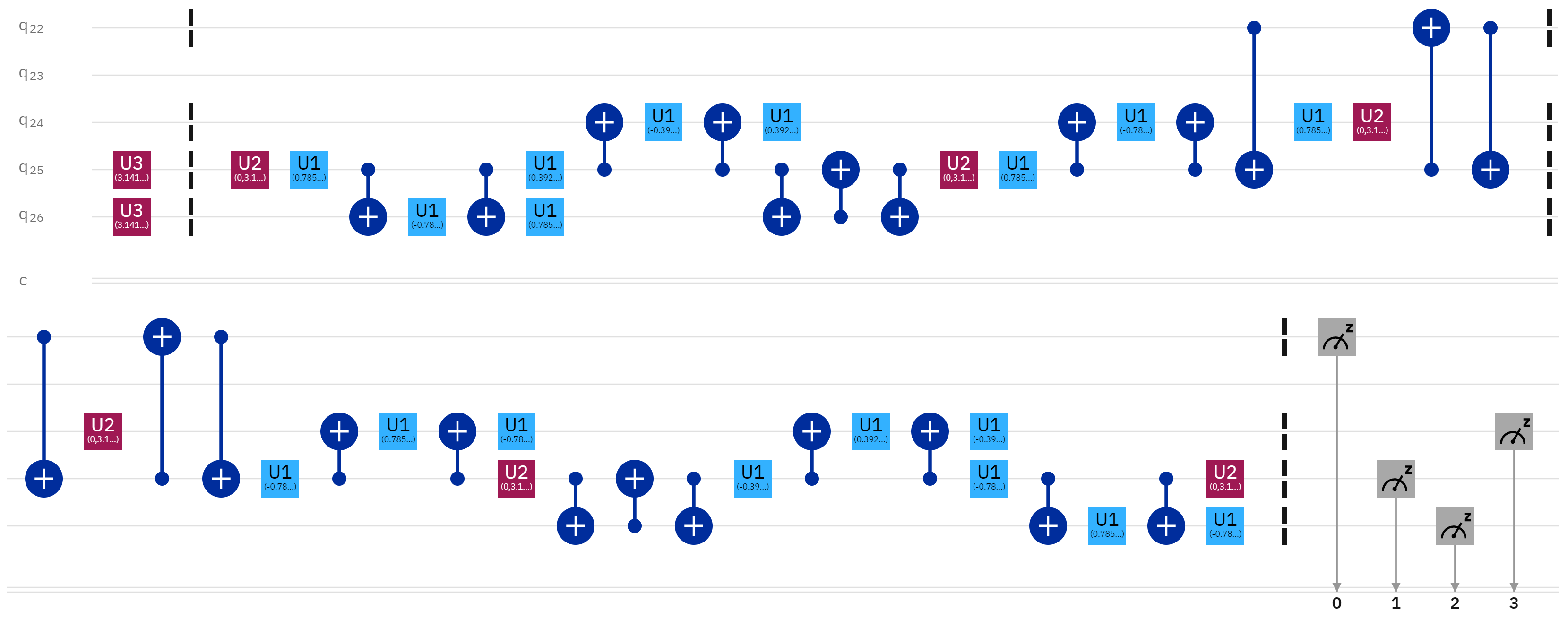}
		\caption{Quantum composition run on IBM's Toronto chip for the 4-qubit QFT experiment (results shown in figure \ref{Fig: QFT Data}), achieving the quantum circuit outlined in figure \ref{Fig: HH QFT} for the initial state $|011\rangle$.}
		\label{Apx: IBMQ QFT}
	\end{figure}
	
	\begin{figure}[h]                     
		\centering
		\includegraphics[scale=.3]{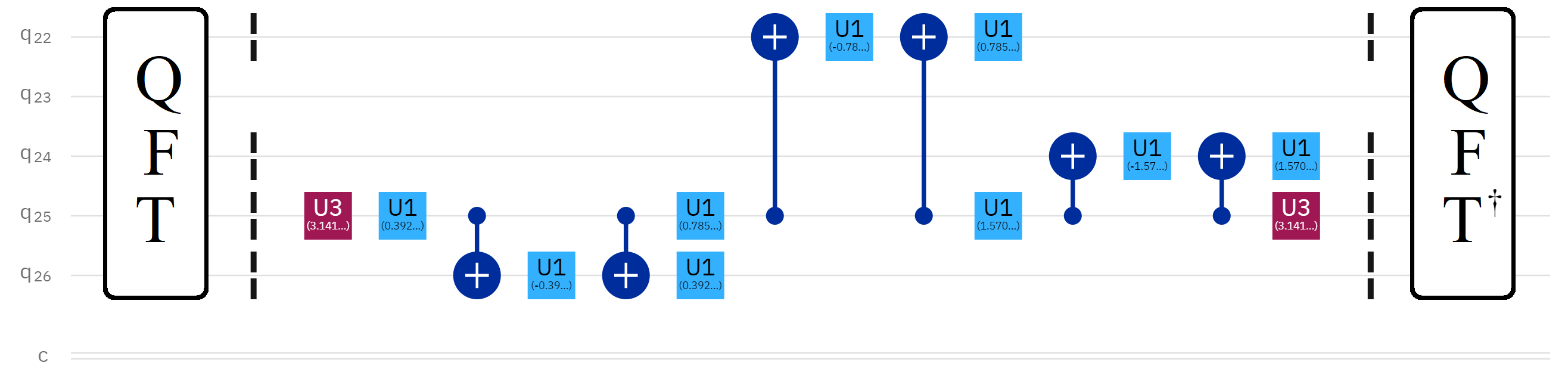}
		\caption{Quantum composition run on IBM's Toronto chip for the 4-qubit Shift Operator experiment (results shown in figure \ref{Fig: Shift Data}).  The circuit compositions for QFT and QFT$^{\dagger}$ are shown in figure \ref{Apx: IBMQ QFT}, such that the gates shown between them represent the QADD$^{+}$ operator acting on three node qubits.}
		\label{Apx: IBMQ Shift}
	\end{figure}

\end{document}